\documentclass[
	nofootinbib,
	preprint,
	noshowpacs,
	noshowkeys,
	floatfix,aps
]{revtex4-1}
\usepackage{amsmath,multirow,amssymb,mathtools}
\usepackage[T1]{fontenc}
\usepackage{verbatim}
\usepackage{float} 
\usepackage{amsfonts}
\usepackage{comment}
\usepackage{verbatim}
\usepackage{braket}
\usepackage{rotating}
\usepackage{dsfont}
\usepackage{xcolor}
\usepackage[justification=raggedright]{caption}
\usepackage{bm}
\usepackage{tikz}
\usetikzlibrary{matrix}
\usepackage{xpatch} 
\makeatletter
	\xpatchcmd{\@ssect@ltx}{\@xsect}{\edef\@currentlabelname{#8}\@xsect}{}{}
	\xpatchcmd{\@sect@ltx}{\@xsect}{\edef\@currentlabelname{#8}\@xsect}{}{}
\makeatother
\usepackage[
	pdfstartview=FitBV,
	bookmarks=true,
	bookmarksopen=true,
	colorlinks =true,
 	linkbordercolor=blue,
 	linkcolor = blue,
	citecolor = blue,
	urlcolor=blue 
]{hyperref} 
\newcommand{\ownref}[2]{\hyperref[#2]{#1~\ref*{#2}}}
\usepackage{cleveref}
\graphicspath{ {./Figures/} }

\begin{document}
\title{Quantum dynamics of a polar rotor acted upon by an electric rectangular pulse of variable duration}

\author{Mallikarjun Karra}
\affiliation{
	Fritz-Haber-Institut der Max-Planck-Gesellschaft \\ Faradayweg 4-6, D-14195 Berlin, Germany}
\author{Burkhard Schmidt}
\email{burkhard.schmidt@fu-berlin.de}
\affiliation{
	Institut f\"ur Mathematik, Freie Universit\"{a}t Berlin \\ Arnimallee 6, D-14195 Berlin, Germany}
\author{Bretislav Friedrich}
\email{bretislav.friedrich@fhi-berlin.mpg.de}
\affiliation{
	Fritz-Haber-Institut der Max-Planck-Gesellschaft \\ Faradayweg 4-6, D-14195 Berlin, Germany}

\date{\today}

\begin{abstract}
As demonstrated in our previous work [J. Chem. Phys. \textbf{149}, 174109 (2018)], the kinetic energy imparted to a quantum rotor by a non-resonant electromagnetic pulse with a Gaussian temporal profile exhibits quasi-periodic drops as a function of the pulse duration. Herein, we show that this behaviour can be reproduced with a simple waveform, namely a rectangular electric pulse of variable duration, and examine, both numerically and analytically, its causes. Our analysis reveals that the drops result from the oscillating populations that make up the wavepacket created by the pulse and that they are necessarily accompanied by drops in the orientation and by a restoration of the pre-pulse alignment of the rotor. Handy analytic formulae are derived that allow to predict the pulse durations leading to diminished kinetic energy transfer and orientation. Experimental scenarios are discussed where the phenomenon could be utilized or be detrimental.
\end{abstract}
\maketitle
\section{Introduction}
The ability to manipulate the quantum states of molecular rotors \citep{Lemeshko_Rev2013, Koch_Lemeshko_SugnyRev2019} has found diverse applications ranging from stereodynamics \citep{Stereoselectivity1, Stereoselectivity2, Stereoselectivity3, deMiranda_Bio} to enantioselectivity \citep{Eibenberger2017, Schnell2017, KochEnantioselectivity2019} to coherent control of rotational states \citep{OhshimaHasegawa2010, RotControlAverbukh2002, Leibscher2004, Daems2005, AlignmentHHG2007, RovibronicHyperfine} to quantum information processing \citep{Quantum_Info0, Quantum_Info3, Quantum_Info2, Quantum_Info3,  Quantum_Info1, Quantum_Info4}. Moreover, quantum rotor dynamics induced by periodic $\delta$-pulses  has been linked to quantum chaos, dynamical localisation, and Bloch oscillations \citep{Floss_Averbukh1, Floss_Averbukh2, Bitter_Milner1, Bitter_Milner2, UltrashortQCMilner2011}. 
 
Whereas either adiabatic  interactions \citep{PendularNature, PendularStatesPRL, AuzinshStark1992, Friedrich1999, Friedrich1999a, TI4, TI3, TI2, TI1, TI0,  DC_Manipulation2017, DCManipulation2018, DCManipulation_Account2021} or their impulsive, non-adiabatic counterparts \citep{RevSeidemannStapelfeldt, Seideman_revival, Cai2001, Cai_BF_2001, Jauslin2005, EllipticLaserPulses2005, Rouzee2007, FieldFreeDaemsGuerin2007, GuerinRouzee2008, Owschimikow2009, Owschimikow2011,  FleischerAverbukh2012, THzPulseShaping2017, THzSugny2019, Nautiyal2021} have received much attention, interactions with finite-duration pulses have been scarce \citep{Dion2001, Ortigoso_BF_Dynamics, Mirahmadi2018}.
It is the last that are the focus of the present study. 

The impulsive interaction of a rotor  with a $\delta$-pulse  is analytically solvable and represents a benchmark for the behavior of the expansion (or hybridization) coefficients of the rotational states, $J$, that make up the wavepacket created by the pulse. For a purely orienting interaction that arises for a polar rotor acted upon by an electric field, the hybridization coefficents were found \citep{RotControlAverbukh2002, Leibscher2004, Mirahmadi2018} to be proportional to the spherical Bessel functions of the first kind, $\mathcal{J}_J(P)$, where $P$ is the pulse strength, see below. The zeroes of the hybridization coefficients, see Fig. 1(b) of Ref. \citep{Mirahmadi2018}, determine the minima of the post-pulse populations of the various rotational states and are key to optimizing the post-pulse value of a given observable. The figure also illustrates that, typically, many states are hybridized by a $\delta$-pulse. In the same study, we showed that Gaussian pulses of a finite, albeit narrow temporal width much shorter than the rotational period of the rotor, simulate accurately the hybridization effects of the $\delta$-pulses. However, for longer pulses of small to moderate strength, we observed sudden quasi-periodic drops in the kinetic energy imparted to the rotor by the pulse as a function of the pulse duration, see Fig. 4 of Ref. \citep{Mirahmadi2018}. Herein, we show that the effects of such pulses can be modelled by an even simpler waveform, namely a rectangular electric pulse whose amplitude, $\varepsilon$, is constant over the duration, $s$, of the pulse, see Fig. \ref{Fig1}(a). Moreover, we show that the quasi-periodic drops in the imparted kinetic energy are necessarily accompanied by a vanishing orientation and by a restoration of the pre-pulse alignment of the rotor. We identify the origin of the effect in the oscillations in $s$ of the rotor-state populations (hybridization coefficients) that make up the rotational wavepacket (hybrid). Finally, we provide handy analytic formulae expressed in terms of the characteristics of the pulse and the rotor that allow to predict the pulse durations leading to diminished kinetic energy transfer and orientation and discuss experimental scenarios where the phenomenon could be taken advantage of or be detrimental.

\section{A Model Quantum System}
\begin{figure}
\centering
\includegraphics[scale=0.15]{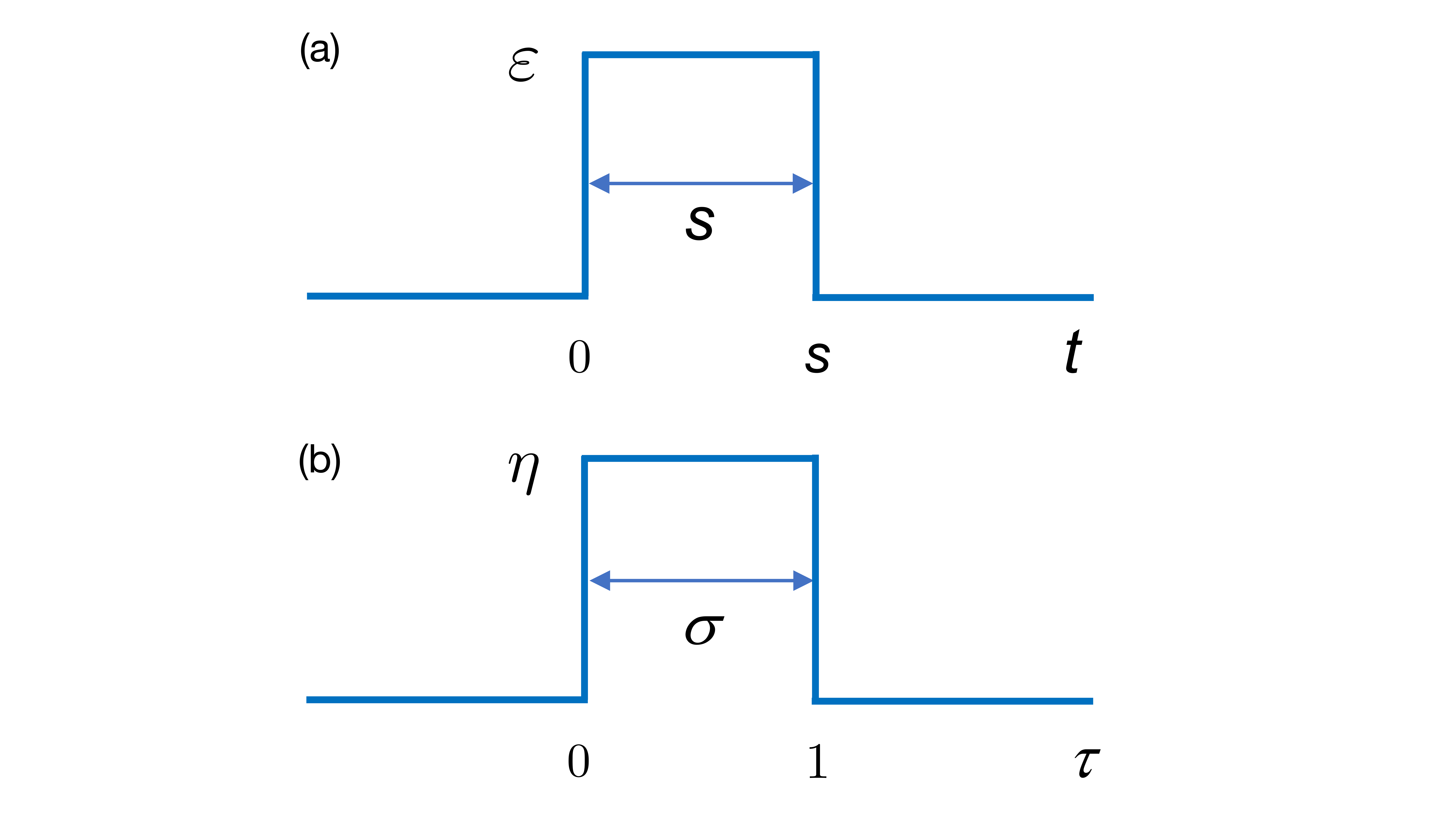}
\caption{A schematic showing the dependence (a) of the electric field strength $\varepsilon$ on time $t$ for a rectangular pulse of duration $s$ and (b) of the orienting parameter $\eta \equiv \mu \varepsilon/B$ on the reduced time $\tau$ for a rectangular pulse of duration $\sigma$.} \label{Fig1}
\end{figure}

We consider a polar linear rigid rotor with angular momentum $\mathbf{J}$ and rotational constant $B = \hbar^2/2I$, where $I$ the moment of inertia, subject to a time-dependent potential
\begin{align}\label{V_genform}
 V(\theta,t) = \left\{
 \begin{tabular}{cr}
 $ -\mu \varepsilon \cos\theta$  & $ 0 \leq t \leq s $ \\
 0  & $ t > s $
 \end{tabular} \right \}
 \end{align}
with $\theta$ the polar angle between the electric field of strength  $\varepsilon$ and the molecular electric dipole moment of magnitude $\mu$. The potential is nonzero only during time $0 \leq t \leq s$ and thus corresponds to a rectangular pulse of amplitude $\mu \varepsilon \cos\theta$ and duration $s$, see also Fig. \ref{Fig1}(a).

The corresponding time-dependent Schr\"odinger equation (TDSE) 
\begin{equation}
 i\hbar \frac{\partial}{\partial t} \ket{\psi(t)} = \big[B\mathbf{J}^2+V(\theta,t)\big] \ket{\psi(t)}\end{equation}
can be recast, upon rescaling time in units of the pulse duration, $\tau\equiv t/s$, as
\begin{equation}\label{tdse_rescaled}
 i \frac{\partial}{\partial \tau} \ket{\psi(\tau)} = \big(\sigma\mathbf{J}^2-\eta\sigma\cos\theta\big) \ket{\psi(\tau)}
\end{equation}
where $\sigma=\frac{Bs}{\hbar}$ is the pulse duration in units of the rotational period, $\hbar/B$, and $\eta\equiv \frac{\mu \varepsilon}{B}$ is a dimensionless measure of the orienting interaction. Note that the potential becomes zero at $\tau>1$. Integration of the orienting interaction over the pulse duration gives the pulse strength (or kick strength) $P = \int_{0}^{s}\frac{\mu E}{\hbar} \ \mathrm{d}t=\frac{\mu E s}{\hbar}=\eta\sigma$, see also Fig. \ref{Fig1}(b). 

The matrix representation of the Hamiltonian in the free rotor basis set is symmetric tri-diagonal, with matrix elements 
\begin{equation}
\bra{J',0}\mathbf{J}^2\ket{J,0} = J(J+1)\delta_{J',J}
\end{equation}
and 
\begin{equation}
\bra{J',0}\cos\theta\ket{J,0}=\sqrt{\frac{J^2}{(2J+1)(2J-1)}}\delta_{J',J-1}+\sqrt{\frac{(J+1)^2}{(2J+3)(2J+1)}}\delta_{J',J+1}
\end{equation}  

We note that well-established perturbative approaches, for example the first-order Magnus expansion, or the  time evolution operator of the time-dependent unitary perturbation theory (TDUPT) as developed by Daems \textit{et. al.} \citep{TDUPT2004}, assume small perturbations and are thus only suitable for treating the $\sigma$-dependent effects of our system for $\sigma \ll \sigma \eta$. Furthermore, the potent iterative super-convergent KAM algorithm technique \citep{Superconvergent_KAM1, Superconvergent_KAM2} is tedious to apply beyond the two-level approximation. Therefore, herein we resort to solving the TDSE numerically by the split-operator method using the Gauss-Jacobi quadrature implemented within the WavePacket software package \citep{Schmidt2017, Wavepacket2, Wavepacket3}. Apart from that, we set up a ten-level approximation in Mathematica \citep{Mathematica} and discuss in detail the two-level approximation valid in the limit of small $P$. 

\section{Results and Discussion}\label{Results}

\subsection{Numerical simulations}
The numerical simulations presented in this Section have been carried out as a function of the pulse duration $\sigma$ for a fixed value of the pulse strength $P=\sigma \eta=1.5$. The range of $\sigma$ was varied between $0.005$ and $10$ in steps of $0.005$, i.e., from the impulsive, non-adiabatic regime (the results for $\sigma=0.005$ are in agreement with the theory of $\delta$-kicks \citep{Mirahmadi2018}) to  the adiabatic limit (the results for $\sigma=10$ approximate well the stationary solutions of the TDSE, \cref{tdse_rescaled}). Results for other values of $P$ are presented and discussed in Sec.~\ref{two_ten}.

\begin{figure}
\includegraphics[scale=0.45]{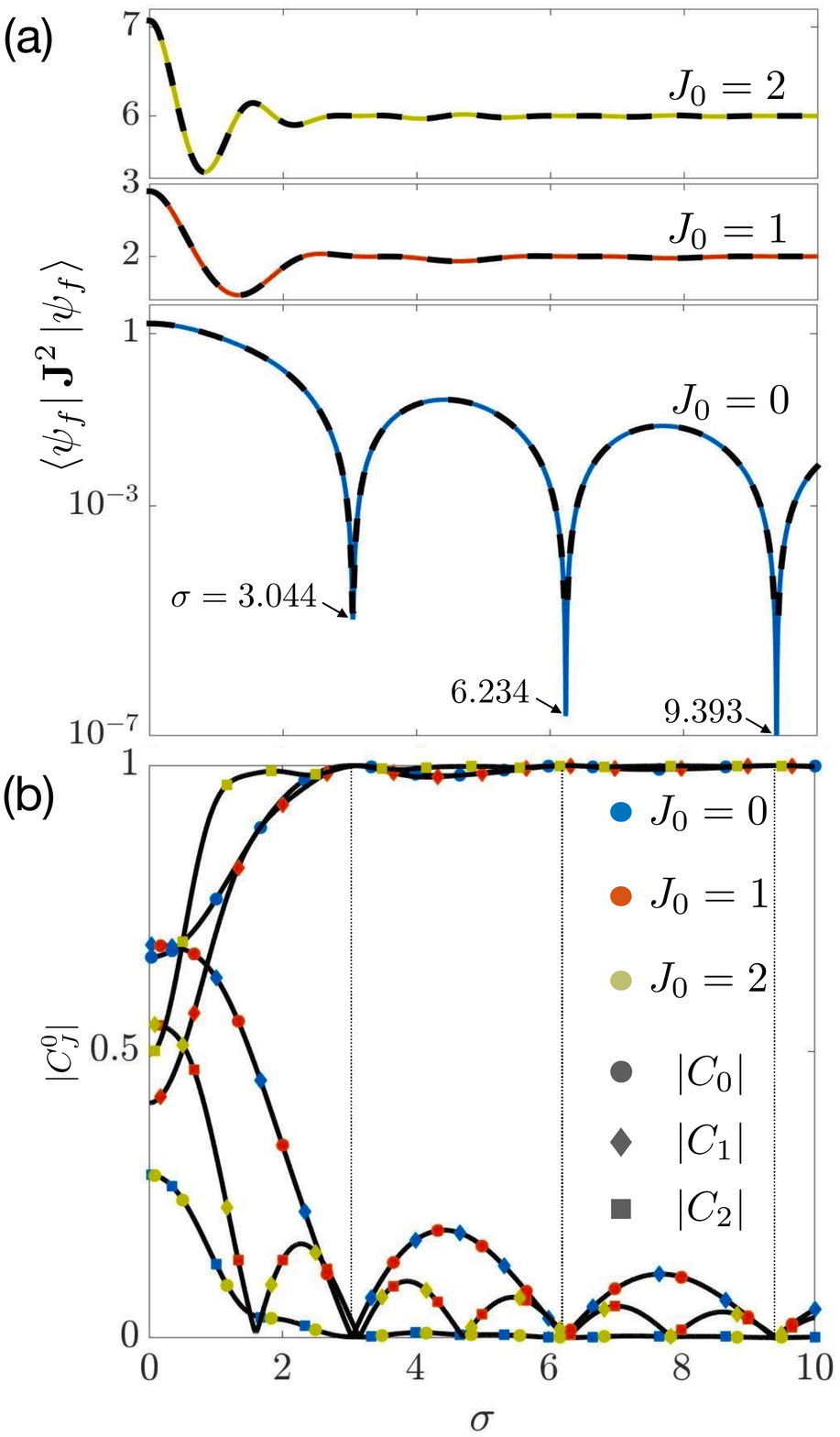}
\caption{(a) Rotational kinetic energy imparted to a polar rigid rotor in an initial state $\ket{J_0, 0}$ with $J_0=0$ (blue), $J_0=1$ (red), $J_0=2$ (yellow) by a rectangular electric pulse as a function of the pulse duration $\sigma$ at a fixed pulse strength $P=1.5$. The dashed black curves represent the kinetic energies obtained from a ten-level approximation presented in Sec.~\ref{two_ten}. (b) Absolute values of the first three expansion coefficients for each of the three initial states. Note the relationship, $C^{i}_{j}=C_{i}^{j}$.} \label{Fig2}
\end{figure}

\begin{figure}
\centering
\includegraphics[scale=0.25]{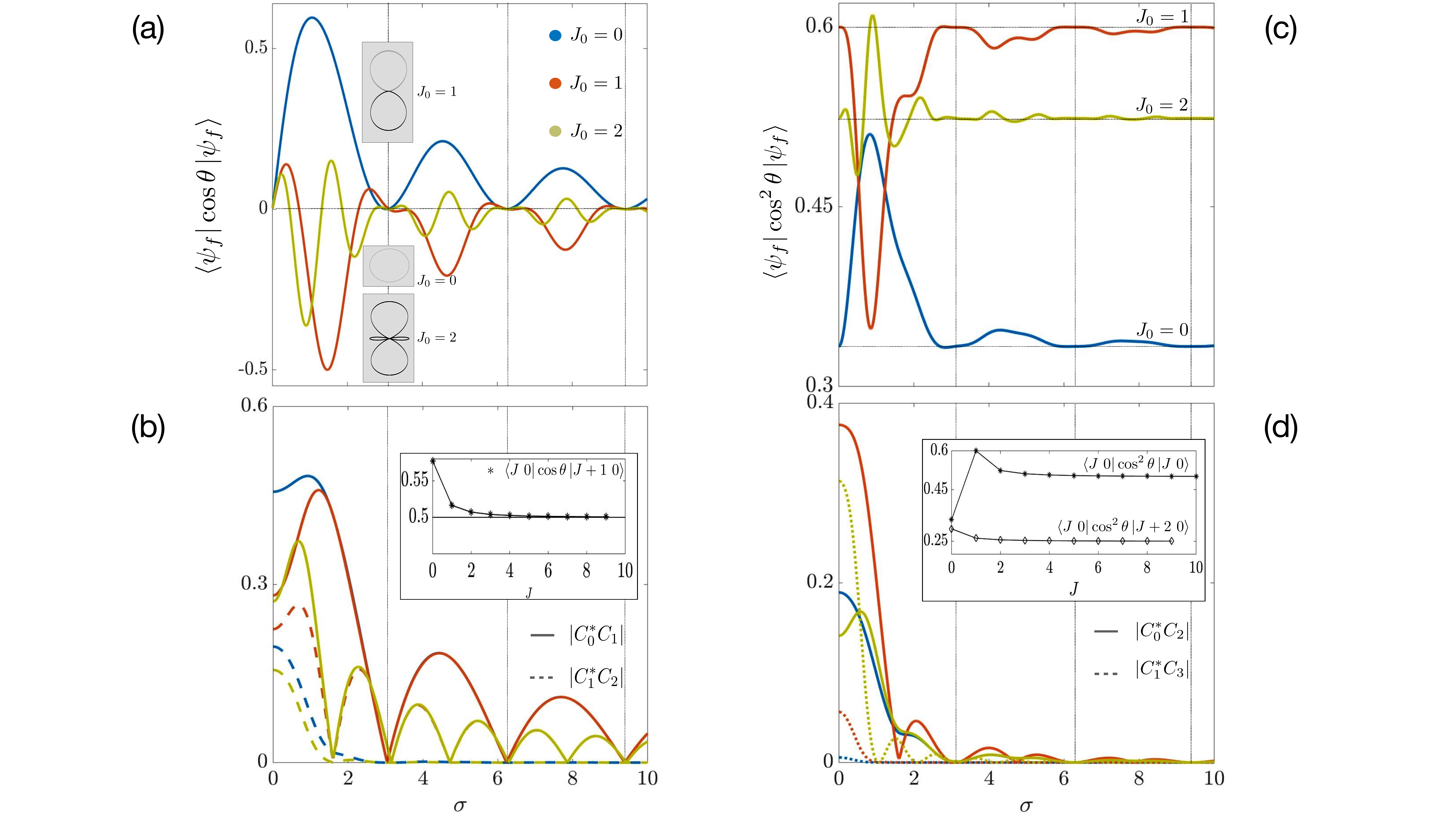}
\caption{Directional properties of the wavepackets created by a rectangular pulse of pulse strength $P=1.5$ as a function of the pulse duration $\sigma$. (a) Post-pulse orientation cosine for initial rotational states with $J_0=0$ (blue), $J_0=1$ (red), and $J_0=2$ (yellow). The insets show polar plots of the wavefunctions at $\sigma=3.044$. (b) Absolute values of the products of the coefficients of successive rotational levels. Inset shows the matrix elements of $\cos\theta$ for $\Delta J = \pm 1$, with the first two values evaluating to $\frac{1}{\sqrt{3}}$ and $\frac{2}{\sqrt{15}}$, respectively, such that the large-$J$ limit tends to 0.5. (c) Post-pulse alignment cosine for initial rotational states with $J_0=0$ (blue), $J_0=1$ (red), and $J_0=2$ (yellow). For the case of a large $\sigma$ (small $\eta$), the alignment returns to its initial value. Horizontal lines show the alignment of the field-free states. (d) Absolute values of the products of the coefficients of successive rotational levels with a difference in $J$ of two. Inset shows the matrix elements of $\cos^2\theta$ for $\Delta J = 0, \ \pm 2$.} \label{Fig3}
\end{figure}

The results for the initial states $\ket{J_0, 0}$ with $J_0 \in \{0,1,2\}$ of the rotor are summarised in Figs.~\ref{Fig2} and \ref{Fig3}. The rotational kinetic energy and the orientation and alignment cosines shown pertain to the state $\ket{\psi_f}$ of the system at the end of the interaction at $\tau=1$ with a pulse of duration $\sigma$ and  orienting interaction parameter $\eta = \frac{P}{\sigma}$. Expanded in the free rotor basis $\ket{J, 0}$, the post-pulse state is given by
\begin{equation}
\ket{\psi_f}=\sum_J C^{J_0}_{J}\ket{J, 0}
\end{equation}
where the superscript of the expansion (hybridization) coefficients, $J_0$, denotes the initial rotational state of the rotor.  

Thus the rotational kinetic energy at the end of the pulse becomes
\begin{equation}
\bra{\psi_f}\mathbf{J}^2\ket{\psi_f} = \sum_J J(J+1) |C_J|^2
\end{equation}
where we dropped the superscript $J_0$ on $C_J$ for notational simplicity. We note that, by symmetry,  $C^{n}_m=C^m_n$.
The dependence of the imparted rotational kinetic energy on the pulse duration, Fig.~\ref{Fig2}(a), exhibits drops that are particularly pronounced for the ground state of the rotor, with $J_0= 0$, but also appear for the higher initial states of the rotor with  $J_0= 1$ and $J_0= 2$. Fig.~\ref{Fig2}(b) shows the dependence of the corresponding hybridization coefficients on the pulse duration. Their squares, $|C_J|^2$, give the populations of the free rotor basis states in a given wavepacket (hybrid). Close to the adiabatic limit (at $\sigma=10$), essentially only the initial states are populated after the pulse has passed, i.e., the state of the rotor before and after an adiabatic interaction is the same. This is in sharp contrast with what happens in the impulsive, non-adiabatic regime at short pulses (small $\sigma$): a wavepacket is created that is comprised of a number of free-rotor states $J$ whose relative contributions oscillate with a $J$-dependent frequency as a function of $\sigma$. The collective vanishing  of the populations of the contributing $J$ states at particular values of $\sigma$ is thus the apparent cause of the drops in the rotational kinetic energy, cf. Figs.~\ref{Fig2}(a) and \ref{Fig2}(b).  The numerically determined drops occur at $\sigma \approx 3.044, 6.234, 9.393$, with a period of a little less than $\pi$ which, as we will see in Sec.~\ref{two_ten}, is due to our choice of $P$. 

We note that for the initial state $J_0=0$, the kinetic energy ($2|C_1|^2$) is minimised when the real part of the $C_1$ coefficient vanishes, i.e., at the three points mentioned above (within the two-state approximation, it does so necessarily due to the sinc prefactor). However, a small imaginary part remains and therefore the kinetic energy does not completely vanish in a large-enough basis. In the case of a weak pulse strength ($P \gtrsim 3$), the $C_J$ coefficients with $J \gtrsim 3$ are found to be negligible for $\sigma \gtrsim 2$.  

For higher initial states ($J_0=1$ and $J_0=2$), the rotational kinetic energy, as a cumulative linear combination of the squares of the coefficients of all populated rotational levels, is found to oscillate in $\sigma$ with a decaying amplitude from its starting value, $\frac{2P^2}{3}$ (valid for $J_0=0$), imparted by an ultrashort pulse  \cite{Mirahmadi2018} to the long-pulse adiabatic limit, $J_0(J_0+1)$, where only the initial state is populated after the pulse has passed, cf. the upper two curves in Fig.~\ref{Fig2}(a). 

The directional properties -- the post-pulse orientation and alignment cosines -- of the wavepacket created by the rectangular pulse are shown in panels (a) and (c) of Fig. \ref{Fig3} as functions of pulse duration $\sigma$ for a fixed pulse strength $P=1.5$.  

The post-pulse orientation cosine of such a wavepacket is given by 
\begin{equation}\label{orientation}
\bra{\psi_f}\cos\theta\ket{\psi_f}=\sum_J C_{J}^*C_{J+1}\bra{J, 0}\cos\theta\ket{J+1, 0}+\mathrm{c.c.}
\end{equation}
where $\ket{\psi_f}$ denotes the final state at the end of the interaction. Owing to the selection rules, the only non-vanishing terms in the expansion are those corresponding to $\Delta J=\pm1$. As seen in Fig.~\ref{Fig3}(a), the orientation cosine vanishes at the values of $\sigma$ that lead to the drops of the rotational kinetic energy. This behavior is a consequence of the vanishing of the products of pairs of successive coefficients, see Fig.~\ref{Fig3}(b), that appear in the expression (\ref{orientation}) for the orientation cosine. The dotted vertical lines emphasize that the vanishing of the orientation cosine and of the successive pairs of the hybridization coefficients indeed occur at the same values of $\sigma$. This behavior is to be expected for any initial state in regions of the $P-\sigma$ space where not too-many free rotor states are hybridized (see discussion pertaining to Fig.~\ref{Fig4}).

The vanishing of the orientation is visualized in Fig.~\ref{Fig3}(a) by three insets representing the polar plots of the wavefunction for the three different initial states at $\sigma=3.044$ (the electric field of the pulse is oriented vertically). The near-symmetry of the lobes about the horizontal axis indicates a negligible orientation cosine. However, for the $J_0=1$ and $J_0=2$ states, the alignment is clearly quite pronounced.

The post-pulse alignment is given by
\begin{equation}
\bra{\psi_f}\cos^2\theta\ket{\psi_f}=\sum_{J}|C_J|^2\bra{J, 0}\cos\theta\ket{J, 0}) + \sum_{J}C_{J}^*C_{J+2}\bra{J, 0}\cos\theta\ket{J+2, 0}+\mathrm{c.c.}
\end{equation}
with nonvanishing matrix elements arising for $\Delta J=0, \pm 2$. Fig.~\ref{Fig3}(c)  shows the alignment cosine as a function of the pulse duration $\sigma$. 

The pairwise vanishing of the products of the hybridization coefficients pertaining to the free rotor states that differ by $\Delta J=\pm2$ restores the alignment the rotor had in its initial state $J_0$. 
Fig.~\ref{Fig3}(c) demonstrates that particular values of the pulse duration can maximise the alignment. Its degree depends, however, on the pulse strength. 

\subsection{Two-level approximation}\label{two_ten}
Within the first-order perturbation theory, the probability $P_{mk}$ of a transition from state $\ket{m}$ to state $\ket{n}$ of a two-level quantum system driven by a constant perturbation $V$ is given by $P_{mk} = \frac{V^2t^2}{\hbar}\mathrm{sinc}^2(t\Delta/2\hbar)$, where $\Delta$ is the energy difference between the two states \citep{cohen1977, Sakurai1994}; the sinc function is defined as $\mathrm{sinc(x)}\equiv \frac{\sin x}{x}$. With this in mind, we re-cast eq.~(\ref{tdse_rescaled}) explicitly for $N$ rotational levels,
\begin{align}\label{matrixeqn}
\begin{pmatrix} \dot{C}_0(\tau)\\
\dot{C}_{1}(\tau) \\
\dot{C}_2(\tau) \\
\cdot \\
\cdot \end{pmatrix} = i\sigma\begin{pmatrix} 0 & \frac{\eta}{\sqrt{3}} & 0 & 0 & \ 0 \\
\frac{\eta}{\sqrt{3}} & -2  & \frac{2\eta}{\sqrt{15}} & \ 0 \ & \ 0 \\
0 & \frac{2\eta}{\sqrt{15}} & -6 & \  \cdot \ & \ 0 \\
0 & 0 & \cdot & \ \cdot  \ &  \ \cdot \\
0 & 0 & 0 & \cdot &  \ \cdot \end{pmatrix} \begin{pmatrix} C_0(\tau)\\
C_1(\tau) \\
C_2(\tau) \\
\cdot \\
\cdot \end{pmatrix}
\end{align}
where the left-hand side is a vector of time derivatives  of the expansion coefficients with respect to dimensionless time $\tau$ of the $N$ rotational levels.

We now consider the following two-level models formed out of two successive $2\times2$ diagonal sub-blocks of the matrix in eq. (\ref{matrixeqn}). A model constructed from the $\ket{0,0}$ and $\ket{1,0}$ states,
\begin{align}\label{J_00}
i\sigma\begin{pmatrix}
0 & \frac{\eta}{\sqrt{3}} \\
\frac{\eta}{\sqrt{3}}  & -2 \end{pmatrix} \hspace{2cm} \mathrm{for}\ J_0=0
\end{align}
 and a model constructed from the $\ket{1,0}$ and $\ket{2,0}$ states,
 \begin{align}\label{J_01}
 i\sigma\begin{pmatrix}
-2 & \frac{2\eta}{\sqrt{15}} \\
\frac{2\eta}{\sqrt{15}}  & -6 \end{pmatrix} \hspace{2cm} \mathrm{for}\ J_0=1
\end{align}
The general solutions for the expansion coefficients constructed from the eigenvalues, $\lambda$, and eigenvectors, $\hat{\pmb{\nu}}$, of the matrices are given by the following equation:
\begin{equation}\label{C01}
C^{J_0}(\tau)=\mathcal{A}^{J_0}_1e^{\lambda^{J_0}_1\tau}\hat{\bm{\nu}}^{J_0}_{1} +\mathcal{A}^{J_0}_2e^{\lambda^{J_0}_2\tau}\hat{\bm{\nu}}^{J_0}_{2}  
\end{equation}
where $\mathcal{A}^{J_0}_{1,2}$ are the constants of integration obtained by imposing the initial conditions ($J_0=0$ and $J_0=1$, respectively). Table~\ref{Table1} summarizes the eigenvalues and eigenvectors of the two corresponding block-diagonal sub-matrices. 

By substituting $\tau=1$ into eq. (\ref{C01}), models (\ref{J_00}) and (\ref{J_01}) render, respectively, the following expressions for the coefficients of the initially unpopulated states, 
\begin{equation}\label{C0}
C_{1}^{0}(\sigma, P) = \frac{i P \ \mathrm{sinc}(\sigma\xi^{0})}{\sqrt{3}}\exp(i \sigma)
\end{equation}
\begin{equation}\label{C1}
C_{2}^{1}(\sigma, P) = \frac{2i P \ \mathrm{sinc}(\sigma\xi^{1})}{\sqrt{15}}\exp(4i  \sigma)
\end{equation}
where the argument of the sinc function, $\xi^{J_0}$, is defined in Table~\ref{Table1}; it is proportional to the difference of the two eigenvalues -- and thus analogous to the factor $\frac{t\Delta}{2\hbar}$ appearing in the perturbative treatment mentioned above. The zeroes of the sinc function (which coincides with the zeroth-order spherical Bessel function of the first kind, $\mathcal{J}_0$) occur at integer multiples of $\pi$ of the argument $\xi$. Hence for integer $n$ and a  pulse strength $P$, the real-valued roots of the equations 
\begin{equation}\label{sig1}
\sigma^0_n = \sqrt{\frac{3n^2\pi^2-P^2}{3}} \approx n\pi\bigg(1-\frac{P^2}{6n^2\pi^2}\bigg)
\end{equation}
\begin{equation}\label{sig2}
\sigma^1_n = \sqrt{\frac{15n^2\pi^2-4P^2}{60}} \approx \frac{n\pi}{2}\bigg(1-\frac{2P^2}{15n^2\pi^2}\bigg)
\end{equation}
yield, respectively, the values of the pulse durations at which the $C_1^0$ and $C_2^1$ coefficients vanish. The first-order Taylor expansions of these pulse durations, included in eqs. (\ref{sig1}) and (\ref{sig2}), indicate that the dependence on $P$ of the loci of the zeros of the $C_1^0$ and $C_2^1$ coefficients is  parabolic. In the weak-perturbation limit, $P\to 0$, the coefficients $C^0_1 (=C^1_0)$ and $C^1_2 (=C^2_1)$ have periods $\pi$ and $\pi/2$, respectively. 

\begin{table}[h!]
  \begin{center}
    \begin{tabular}{r|c|c} 
       &  $\ket{00}$ and $\ket{10}$ & $\ket{10}$ and $\ket{20}$ \\
       & $J_0=0$ & $J_0=1$ \\
      \hline
      Eigenvalues ($\lambda^{J_0}_{1,2}$)  & $\lambda^0_{1,2}=-i \sigma\bigg(1\pm\sqrt{1+\frac{\eta^2}{3}}\bigg)$ & $\lambda^1_{1,2}=-2i \sigma\bigg(2\pm\sqrt{1+\frac{\eta^2}{15}}\bigg)$ \\
      Eigenvectors ($\hat{\bm{\nu}}^{J_0}_{1,2}$) & $\bigg\{\frac{\sqrt{3}}{\eta}\bigg(1\mp\sqrt{1+\frac{\eta^2}{3}}\bigg),1 \bigg\}$  &  $\bigg\{\frac{\sqrt{15}}{\eta}\bigg(1\mp\sqrt{1+\frac{\eta^2}{15}}\bigg),1 \bigg\}$\\ 
      $\mathcal{A}^{J_0}_{1,2}$ & $\mathcal{A}^{0}_{1}=-\mathcal{A}^{0}_{2}=-\frac{\eta}{2\sqrt{3+\eta^2}}$ & $\mathcal{A}^{1}_{1}=-\mathcal{A}^{1}_{2}=-\frac{\eta}{2\sqrt{15+\eta^2}}$ \\
      $\xi^{J_0}$ & $\xi^0=\sqrt{1+\frac{\eta^2}{3}}$ & $\xi^1=2\sqrt{1+\frac{\eta^2}{15}}$\\
      \end{tabular}
    \caption{Summary of the eigenproperties of the $2\times2$ Hamiltonian matrices with two different initial populations.  Here $\xi^{J_0}$ is proportional to the argument of the sinc function in eq.~(\ref{C0}) and (\ref{C1}) and $\mathcal{A}^{J_0}_{1,2}$ are the initial-state dependent constants of integration.}
    \label{Table1}
  \end{center}
\end{table}

\begin{figure}
\includegraphics[scale=0.25]{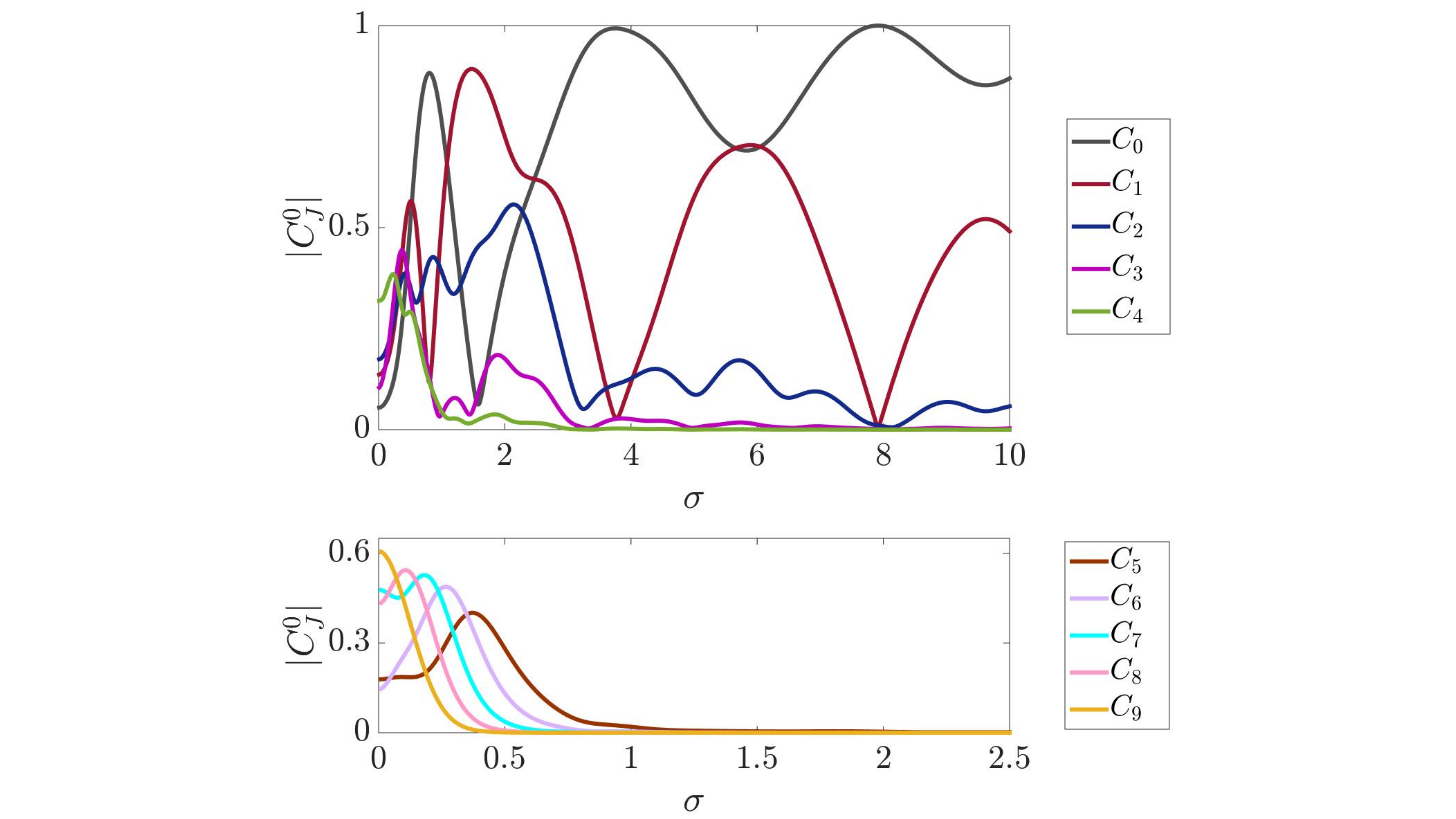}
\centering
\caption{Expansion coefficients for $P=10$ as a function of the pulse-duration $\sigma$ as obtained from the ten-level calculation. There is  gradual transition with increasing $\sigma$ to the regime where the two-level model is approximately valid (as seen in the `bouncing' of the $C_1$ coefficient).} \label{Fig4}
\end{figure}

Furthermore, eqs. (\ref{sig1}) and (\ref{sig2}) show that the coefficients $C^0_1$ and $C^1_2$ vanish for all positive integer values of $n$ provided the pulse strength satisfies the condition $P<\sqrt{3}\pi$ and  $P<\frac{\sqrt{15}}{2}\pi$, respectively.  For larger values of $P$, we demand
$n \geq \frac{P}{\sqrt{3}\pi}$
in order for the solutions $\sigma$ of eq. (\ref{sig1}) to be real-valued. Hence when increasing $P$, the solutions for low $n$ will disappear one after another. For example, for $P=10$, this means that $n \geq 2$, as illustrated in Fig. \ref{Fig4}. A similar argument applies for the solutions of eq. (\ref{sig2}).

For integer values of $n=\{1,2,3\}$, eq.~(\ref{sig1}) yields the loci of vanishing $C^0_1$  at $\sigma_1^0=\{3.022,6.224,9.384\}$ for $P=1.5$, in close agreement with the values obtained numerically, cf. Fig. \ref{Fig2}(b). Likewise, from eq.~(\ref{sig2}) with integer values of $n=\{1,2,3,4\}$, we obtain the loci of the minima of the coefficient $C^1_2$ at $\sigma_2^1=\{1.523,3.113,4.693,6.269\}$ for $P=1.5$, likewise in good agreement with those seen in Fig.~\ref{Fig2}(b). 

\begin{figure}
\includegraphics[scale=0.25]{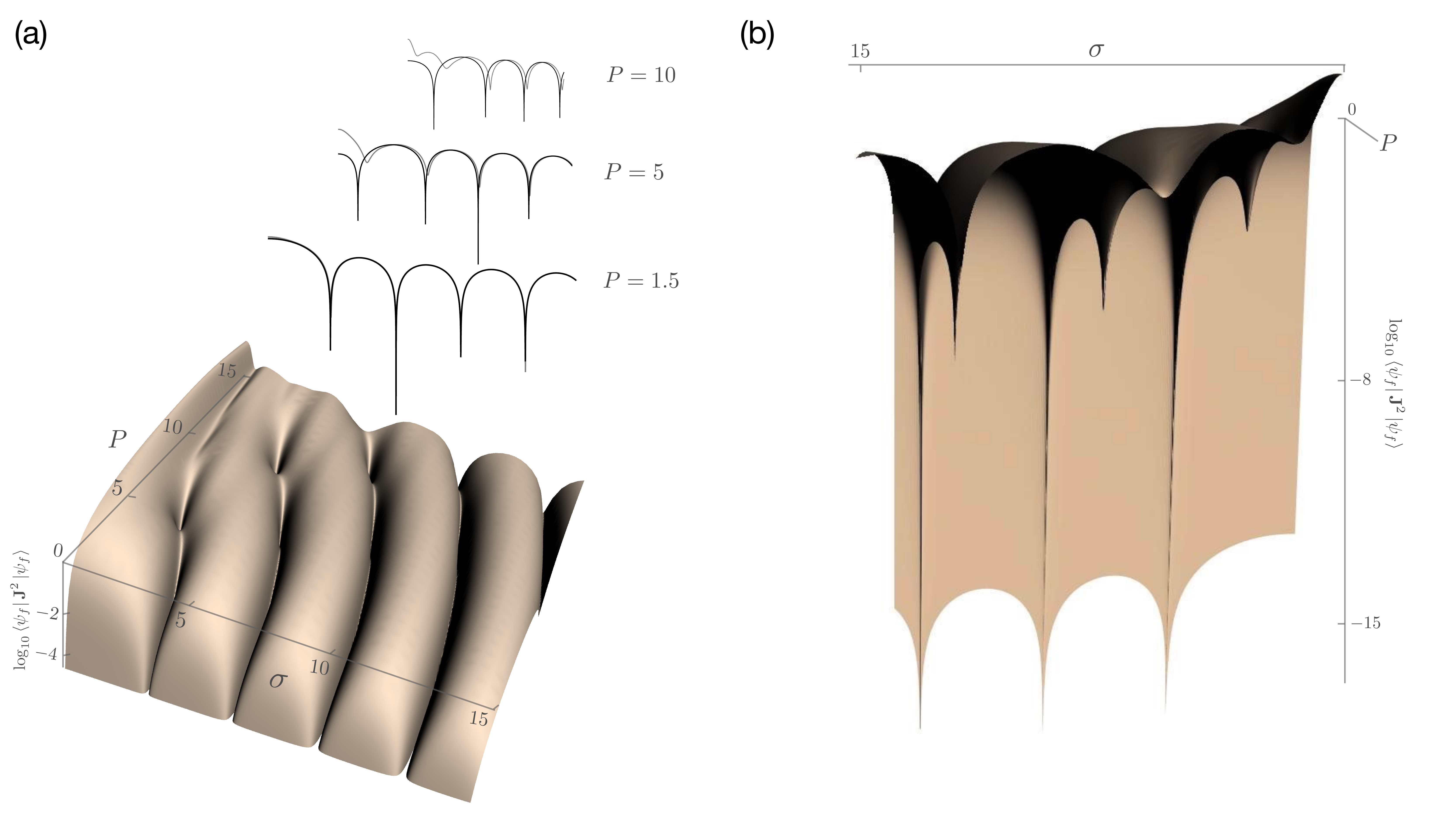}
\centering
\caption{3D views of the $\mathrm{Log}_{10}$-scaled post-pulse kinetic energy surface for the rotor initially in its ground state. Panel (a) is complemented by a comparison of the calculations of the kinetic energy versus pulse duration curves for fixed values of the pulse strength $P$ as obtained from the two-state model (black) and an accurate numerical calculation (grey). Panel (b) provides a view of the minima of the rotational kinetic energy surface.} \label{Fig5}
\end{figure}

Fig.~\ref{Fig4} also illustrates that the behavior of the expansion coefficients is quite different for stronger perturbations ($P \gtrsim 3$ and small enough values of $\sigma$) than for weaker ones. In the short-pulse limit,  both the interaction strength $\eta=P/\sigma$ and the nonadiabaticity increase, leading to the hybridization of higher rotational states. However, their coefficients vanish quicker with increasing $\sigma$ than those of the lower states. As a result, in this $\sigma$ regime, the two- or three-level approximation is approximately valid despite a high value of the pulse strength $P$. 

3D views of the post-pulse rotational kinetic energy surface spanned by the pulse parameters $P$ and $\sigma$ are shown in Fig. \ref{Fig5} for a rotor initially prepared in the ground state $\ket{0,0}$.  Panel (a) shows parabolic furrows, cf. eq.~(\ref{sig1}), consisting of rotationally cooled states with discrete minima along the furrows for specific values of $P$ and $\sigma$, cf. Fig.~\ref{Fig5}(b). These discrete minima occur as a result of the coincidence or near-coincidence of the zeroes of multiple coefficients pertaining to different free-rotor states and correspond to maximally cooled hybrid states arising for particular $P$ and $\sigma$ values. In the absence of closed-form solutions for these coefficients, it is difficult to predict the positions of the minima. However, for the parameter space considered, we find the minima to lie on a family of parallel lines with a slope of $0.577$ in the $P$-$\sigma$ plane. Therefore, given the position of a minimum, one can predict the approximate location of the other minima from the points of intersection of a given line with the family of parabolae described by eq.~(\ref{sig1}). Also shown in Fig. \ref{Fig5}(a) is a comparison of the kinetic energy versus pulse duration curves for fixed values of the pulse strength $P$ as obtained from the two-state model and from an accurate numerical calculation. One can see that for $P < 5$, the two state-model renders the kinetic energy drops quantitatively. The prediction of the two-state model deteriorates at larger $P$ and small $\sigma$ due the hybridization of higher states. Nevertheless, the positions of the drops can still be predicted fairly well for sufficiently large $\sigma$.

\section{Conclusions and Prospects}
In this work, we have identified the reasons behind the sudden quasi-periodic drops of the kinetic energy of a polar rotor subject to a rectangular electric pulse that occur at specific values of the pulse parameters -- its strength $P$ and duration $\sigma$. As a corollary, we found that the orientation of the rotor vanishes at the same values of the pulse parameters as the kinetic energy does, independent of the initial state.

Our study has demonstrated that for not-too-high a value of the pulse strength, it is indeed possible to identify a pulse duration at which the expansion (hybridization) coefficients of the free-rotor states that make up the wavepacket created by the pulse are restricted to just the first few rotational states. An analytic time-dependent two-state model then allowed us to establish a causal connection between the zeros of the hybridization coefficients, as predicted by the model, and the vanishing of the kinetic energy and orientation. 

Current technology makes switching of electrostatic fields at the nanosecond-timescale feasible \cite{meijer_review2012,ns_Switching2,ns_Switching1}. Thus rectangular electric pulses of a nanosecond duration could be generated and used to control large, slowly rotating molecules with a rotational period of a few tens of nanoseconds. By choosing a proper combination of the pulse strength and pulse duration,  isotropic, `un-oriented'  wavepackets with vanishing rotational kinetic energy could be created wherever desirable. In this context, for instance, cold chemistry experiments discounting stereoselective enhancement of reaction rates come to mind. Equally importantly, when undesirable, the phenomenon could also be avoided by a proper tuning of the pulse parameters to ensure that the rotor is properly endowed with both kinetic energy and orientation. 

We hope that an experiment with cold polar linear rotors with small rotational constants ($B\sim0.001$ cm$^{-1}$ and rotational period $\sim 10$ ns) that are subject to rectangular electric pulses will demonstrate the phenomenon studied herein. On the other hand, since we know that full-cycles shorter than the rotational period do indeed result in appreciable changes in orientation (and alignment), smaller rotors with rotational periods in the pico- to femto-second regime could be likewise manipulated by multi-cycle THz pulses \citep{Gao2008, Arkhipov17, Wu2012}. The appropriate parameters for suitably shaping a THz pulse for this purpose could be obtained from our numerical simulations.

Future studies will explore the effect for polar polarizable rotors (i.e., with the aligning term, $\propto \cos^2\theta$, included in the potential) as well as for coherently superposed initial states. Moreover, it would be of interest from a mathematical standpoint to explore the connection  between the quasi-periodic drops in the kinetic energy and directionality, and the reduction of the
error within the second-order TDUPT framework or the iterative super-convergent KAM  
approach, over the $P-\sigma$ parameter space.
A suitable machine-learning algorithm could be employed to `learn' the rules governing low and acceptable error in the TDUPT, and identify islands in the $P$-$\sigma$ parameter space that offer the ability to manipulate a given rotor species. Another avenue to be pursued follows from the observation that a finite-duration pulse is also capable of increasing the kinetic energy (beyond that imparted by an ultrashort non-adiabatic interaction). Amenable to such enhancement of kinetic energy are initial states that are a coherent superposition of several rotational states. This effect could be exploited to generate super-rotors \citep{SuperRotors2021}.

\section*{Acknowledgements}
We thank Gerard Meijer for his interest and support, and Marjan Mirahmadi and Jes\'{u}s P\'{e}rez R\'{i}os for fruitful discussions. MK acknowledges support by the IMPRS for Elementary Processes in Physical Chemistry.

\emph{We dedicate this paper, with much admiration and our best wishes, to J\"urgen Troe on the occasion of his grand jubilee}.

%


\begin{thebibliography}{72}%
\makeatletter
\providecommand \@ifxundefined [1]{%
 \@ifx{#1\undefined}
}%
\providecommand \@ifnum [1]{%
 \ifnum #1\expandafter \@firstoftwo
 \else \expandafter \@secondoftwo
 \fi
}%
\providecommand \@ifx [1]{%
 \ifx #1\expandafter \@firstoftwo
 \else \expandafter \@secondoftwo
 \fi
}%
\providecommand \natexlab [1]{#1}%
\providecommand \enquote  [1]{``#1''}%
\providecommand \bibnamefont  [1]{#1}%
\providecommand \bibfnamefont [1]{#1}%
\providecommand \citenamefont [1]{#1}%
\providecommand \href@noop [0]{\@secondoftwo}%
\providecommand \href [0]{\begingroup \@sanitize@url \@href}%
\providecommand \@href[1]{\@@startlink{#1}\@@href}%
\providecommand \@@href[1]{\endgroup#1\@@endlink}%
\providecommand \@sanitize@url [0]{\catcode `\\12\catcode `\$12\catcode
  `\&12\catcode `\#12\catcode `\^12\catcode `\_12\catcode `\%12\relax}%
\providecommand \@@startlink[1]{}%
\providecommand \@@endlink[0]{}%
\providecommand \url  [0]{\begingroup\@sanitize@url \@url }%
\providecommand \@url [1]{\endgroup\@href {#1}{\urlprefix }}%
\providecommand \urlprefix  [0]{URL }%
\providecommand \Eprint [0]{\href }%
\providecommand \doibase [0]{http://dx.doi.org/}%
\providecommand \selectlanguage [0]{\@gobble}%
\providecommand \bibinfo  [0]{\@secondoftwo}%
\providecommand \bibfield  [0]{\@secondoftwo}%
\providecommand \translation [1]{[#1]}%
\providecommand \BibitemOpen [0]{}%
\providecommand \bibitemStop [0]{}%
\providecommand \bibitemNoStop [0]{.\EOS\space}%
\providecommand \EOS [0]{\spacefactor3000\relax}%
\providecommand \BibitemShut  [1]{\csname bibitem#1\endcsname}%
\let\auto@bib@innerbib\@empty
\bibitem [{\citenamefont {Lemeshko}\ \emph {et~al.}(2013)\citenamefont
  {Lemeshko}, \citenamefont {Krems}, \citenamefont {Doyle},\ and\ \citenamefont
  {Kais}}]{Lemeshko_Rev2013}%
  \BibitemOpen
  \bibfield  {author} {\bibinfo {author} {\bibfnamefont {M.}~\bibnamefont
  {Lemeshko}}, \bibinfo {author} {\bibfnamefont {R.~V.}\ \bibnamefont {Krems}},
  \bibinfo {author} {\bibfnamefont {J.~M.}\ \bibnamefont {Doyle}}, \ and\
  \bibinfo {author} {\bibfnamefont {S.}~\bibnamefont {Kais}},\ }\href {\doibase
  10.1080/00268976.2013.813595} {\bibfield  {journal} {\bibinfo  {journal}
  {Molecular Physics}\ }\textbf {\bibinfo {volume} {111}},\ \bibinfo {pages}
  {1648} (\bibinfo {year} {2013})},\ \Eprint
  {http://arxiv.org/abs/https://doi.org/10.1080/00268976.2013.813595}
  {https://doi.org/10.1080/00268976.2013.813595} \BibitemShut {NoStop}%
\bibitem [{\citenamefont {Koch}\ \emph {et~al.}(2019)\citenamefont {Koch},
  \citenamefont {Lemeshko},\ and\ \citenamefont
  {Sugny}}]{Koch_Lemeshko_SugnyRev2019}%
  \BibitemOpen
  \bibfield  {author} {\bibinfo {author} {\bibfnamefont {C.~P.}\ \bibnamefont
  {Koch}}, \bibinfo {author} {\bibfnamefont {M.}~\bibnamefont {Lemeshko}}, \
  and\ \bibinfo {author} {\bibfnamefont {D.}~\bibnamefont {Sugny}},\ }\href
  {\doibase 10.1103/RevModPhys.91.035005} {\bibfield  {journal} {\bibinfo
  {journal} {Rev. Mod. Phys.}\ }\textbf {\bibinfo {volume} {91}},\ \bibinfo
  {pages} {035005} (\bibinfo {year} {2019})}\BibitemShut {NoStop}%
\bibitem [{\citenamefont {Larsen}\ \emph {et~al.}(2000)\citenamefont {Larsen},
  \citenamefont {Hald}, \citenamefont {Bjerre}, \citenamefont {Stapelfeldt},\
  and\ \citenamefont {Seideman}}]{Stereoselectivity1}%
  \BibitemOpen
  \bibfield  {author} {\bibinfo {author} {\bibfnamefont {J.~J.}\ \bibnamefont
  {Larsen}}, \bibinfo {author} {\bibfnamefont {K.}~\bibnamefont {Hald}},
  \bibinfo {author} {\bibfnamefont {N.}~\bibnamefont {Bjerre}}, \bibinfo
  {author} {\bibfnamefont {H.}~\bibnamefont {Stapelfeldt}}, \ and\ \bibinfo
  {author} {\bibfnamefont {T.}~\bibnamefont {Seideman}},\ }\href {\doibase
  10.1103/PhysRevLett.85.2470} {\bibfield  {journal} {\bibinfo  {journal}
  {Phys. Rev. Lett.}\ }\textbf {\bibinfo {volume} {85}},\ \bibinfo {pages}
  {2470} (\bibinfo {year} {2000})}\BibitemShut {NoStop}%
\bibitem [{\citenamefont {Larsen}\ \emph
  {et~al.}(1999{\natexlab{a}})\citenamefont {Larsen}, \citenamefont
  {Wendt-Larsen},\ and\ \citenamefont {Stapelfeldt}}]{Stereoselectivity2}%
  \BibitemOpen
  \bibfield  {author} {\bibinfo {author} {\bibfnamefont {J.~J.}\ \bibnamefont
  {Larsen}}, \bibinfo {author} {\bibfnamefont {I.}~\bibnamefont
  {Wendt-Larsen}}, \ and\ \bibinfo {author} {\bibfnamefont {H.}~\bibnamefont
  {Stapelfeldt}},\ }\href {\doibase 10.1103/PhysRevLett.83.1123} {\bibfield
  {journal} {\bibinfo  {journal} {Phys. Rev. Lett.}\ }\textbf {\bibinfo
  {volume} {83}},\ \bibinfo {pages} {1123} (\bibinfo {year}
  {1999}{\natexlab{a}})}\BibitemShut {NoStop}%
\bibitem [{\citenamefont {Larsen}\ \emph
  {et~al.}(1999{\natexlab{b}})\citenamefont {Larsen}, \citenamefont {Sakai},
  \citenamefont {Safvan}, \citenamefont {Wendt-Larsen},\ and\ \citenamefont
  {Stapelfeldt}}]{Stereoselectivity3}%
  \BibitemOpen
  \bibfield  {author} {\bibinfo {author} {\bibfnamefont {J.~J.}\ \bibnamefont
  {Larsen}}, \bibinfo {author} {\bibfnamefont {H.}~\bibnamefont {Sakai}},
  \bibinfo {author} {\bibfnamefont {C.~P.}\ \bibnamefont {Safvan}}, \bibinfo
  {author} {\bibfnamefont {I.}~\bibnamefont {Wendt-Larsen}}, \ and\ \bibinfo
  {author} {\bibfnamefont {H.}~\bibnamefont {Stapelfeldt}},\ }\href {\doibase
  10.1063/1.480112} {\bibfield  {journal} {\bibinfo  {journal} {The Journal of
  Chemical Physics}\ }\textbf {\bibinfo {volume} {111}},\ \bibinfo {pages}
  {7774} (\bibinfo {year} {1999}{\natexlab{b}})},\ \Eprint
  {http://arxiv.org/abs/https://doi.org/10.1063/1.480112}
  {https://doi.org/10.1063/1.480112} \BibitemShut {NoStop}%
\bibitem [{\citenamefont {de~Miranda}\ \emph {et~al.}(2011)\citenamefont
  {de~Miranda}, \citenamefont {Chotia}, \citenamefont {Neyenhuis},
  \citenamefont {Wang}, \citenamefont {Qu{\'e}m{\'e}ner}, \citenamefont
  {Ospelkaus}, \citenamefont {Bohn}, \citenamefont {Ye},\ and\ \citenamefont
  {Jin}}]{deMiranda_Bio}%
  \BibitemOpen
  \bibfield  {author} {\bibinfo {author} {\bibfnamefont {M.~H.~G.}\
  \bibnamefont {de~Miranda}}, \bibinfo {author} {\bibfnamefont
  {A.}~\bibnamefont {Chotia}}, \bibinfo {author} {\bibfnamefont
  {B.}~\bibnamefont {Neyenhuis}}, \bibinfo {author} {\bibfnamefont
  {D.}~\bibnamefont {Wang}}, \bibinfo {author} {\bibfnamefont {G.}~\bibnamefont
  {Qu{\'e}m{\'e}ner}}, \bibinfo {author} {\bibfnamefont {S.}~\bibnamefont
  {Ospelkaus}}, \bibinfo {author} {\bibfnamefont {J.~L.}\ \bibnamefont {Bohn}},
  \bibinfo {author} {\bibfnamefont {J.}~\bibnamefont {Ye}}, \ and\ \bibinfo
  {author} {\bibfnamefont {D.~S.}\ \bibnamefont {Jin}},\ }\href {\doibase
  10.1038/nphys1939} {\bibfield  {journal} {\bibinfo  {journal} {Nature
  Physics}\ }\textbf {\bibinfo {volume} {7}},\ \bibinfo {pages} {502} (\bibinfo
  {year} {2011})}\BibitemShut {NoStop}%
\bibitem [{\citenamefont {Eibenberger}\ \emph {et~al.}(2017)\citenamefont
  {Eibenberger}, \citenamefont {Doyle},\ and\ \citenamefont
  {Patterson}}]{Eibenberger2017}%
  \BibitemOpen
  \bibfield  {author} {\bibinfo {author} {\bibfnamefont {S.}~\bibnamefont
  {Eibenberger}}, \bibinfo {author} {\bibfnamefont {J.}~\bibnamefont {Doyle}},
  \ and\ \bibinfo {author} {\bibfnamefont {D.}~\bibnamefont {Patterson}},\
  }\href {\doibase 10.1103/PhysRevLett.118.123002} {\bibfield  {journal}
  {\bibinfo  {journal} {Phys. Rev. Lett.}\ }\textbf {\bibinfo {volume} {118}},\
  \bibinfo {pages} {123002} (\bibinfo {year} {2017})}\BibitemShut {NoStop}%
\bibitem [{\citenamefont {P{\'e}rez}\ \emph {et~al.}(2017)\citenamefont
  {P{\'e}rez}, \citenamefont {Steber}, \citenamefont {Domingos}, \citenamefont
  {Krin}, \citenamefont {Schmitz},\ and\ \citenamefont
  {Schnell}}]{Schnell2017}%
  \BibitemOpen
  \bibfield  {author} {\bibinfo {author} {\bibfnamefont {C.}~\bibnamefont
  {P{\'e}rez}}, \bibinfo {author} {\bibfnamefont {A.~L.}\ \bibnamefont
  {Steber}}, \bibinfo {author} {\bibfnamefont {S.~R.}\ \bibnamefont
  {Domingos}}, \bibinfo {author} {\bibfnamefont {A.}~\bibnamefont {Krin}},
  \bibinfo {author} {\bibfnamefont {D.}~\bibnamefont {Schmitz}}, \ and\
  \bibinfo {author} {\bibfnamefont {M.}~\bibnamefont {Schnell}},\ }\href
  {\doibase https://doi.org/10.1002/anie.201704901} {\bibfield  {journal}
  {\bibinfo  {journal} {Angewandte Chemie International Edition}\ }\textbf
  {\bibinfo {volume} {56}},\ \bibinfo {pages} {12512} (\bibinfo {year}
  {2017})},\ \Eprint
  {http://arxiv.org/abs/https://onlinelibrary.wiley.com/doi/pdf/10.1002/anie.201704901}
  {https://onlinelibrary.wiley.com/doi/pdf/10.1002/anie.201704901} \BibitemShut
  {NoStop}%
\bibitem [{\citenamefont {Leibscher}\ \emph {et~al.}(2019)\citenamefont
  {Leibscher}, \citenamefont {Giesen},\ and\ \citenamefont
  {Koch}}]{KochEnantioselectivity2019}%
  \BibitemOpen
  \bibfield  {author} {\bibinfo {author} {\bibfnamefont {M.}~\bibnamefont
  {Leibscher}}, \bibinfo {author} {\bibfnamefont {T.~F.}\ \bibnamefont
  {Giesen}}, \ and\ \bibinfo {author} {\bibfnamefont {C.~P.}\ \bibnamefont
  {Koch}},\ }\href {\doibase 10.1063/1.5097406} {\bibfield  {journal} {\bibinfo
   {journal} {The Journal of Chemical Physics}\ }\textbf {\bibinfo {volume}
  {151}},\ \bibinfo {pages} {014302} (\bibinfo {year} {2019})},\ \Eprint
  {http://arxiv.org/abs/https://doi.org/10.1063/1.5097406}
  {https://doi.org/10.1063/1.5097406} \BibitemShut {NoStop}%
\bibitem [{\citenamefont {Ohshima}\ and\ \citenamefont
  {Hasegawa}(2010)}]{OhshimaHasegawa2010}%
  \BibitemOpen
  \bibfield  {author} {\bibinfo {author} {\bibfnamefont {Y.}~\bibnamefont
  {Ohshima}}\ and\ \bibinfo {author} {\bibfnamefont {H.}~\bibnamefont
  {Hasegawa}},\ }\href {\doibase 10.1080/0144235X.2010.511769} {\bibfield
  {journal} {\bibinfo  {journal} {International Reviews in Physical Chemistry}\
  }\textbf {\bibinfo {volume} {29}},\ \bibinfo {pages} {619} (\bibinfo {year}
  {2010})},\ \Eprint
  {http://arxiv.org/abs/https://doi.org/10.1080/0144235X.2010.511769}
  {https://doi.org/10.1080/0144235X.2010.511769} \BibitemShut {NoStop}%
\bibitem [{\citenamefont {Averbukh}\ \emph {et~al.}(2003)\citenamefont
  {Averbukh}, \citenamefont {Arvieu},\ and\ \citenamefont
  {Leibscher}}]{RotControlAverbukh2002}%
  \BibitemOpen
  \bibfield  {author} {\bibinfo {author} {\bibfnamefont {I.~S.}\ \bibnamefont
  {Averbukh}}, \bibinfo {author} {\bibfnamefont {R.}~\bibnamefont {Arvieu}}, \
  and\ \bibinfo {author} {\bibfnamefont {M.}~\bibnamefont {Leibscher}},\ }in\
  \href@noop {} {\emph {\bibinfo {booktitle} {Coherence and Quantum Optics
  VIII}}},\ \bibinfo {editor} {edited by\ \bibinfo {editor} {\bibfnamefont
  {N.~P.}\ \bibnamefont {Bigelow}}, \bibinfo {editor} {\bibfnamefont {J.~H.}\
  \bibnamefont {Eberly}}, \bibinfo {editor} {\bibfnamefont {C.~R.}\
  \bibnamefont {Stroud}}, \ and\ \bibinfo {editor} {\bibfnamefont {I.~A.}\
  \bibnamefont {Walmsley}}}\ (\bibinfo  {publisher} {Springer US},\ \bibinfo
  {address} {Boston, MA},\ \bibinfo {year} {2003})\ pp.\ \bibinfo {pages}
  {71--86}\BibitemShut {NoStop}%
\bibitem [{\citenamefont {Leibscher}\ \emph {et~al.}(2004)\citenamefont
  {Leibscher}, \citenamefont {Averbukh}, \citenamefont {Rozmej},\ and\
  \citenamefont {Arvieu}}]{Leibscher2004}%
  \BibitemOpen
  \bibfield  {author} {\bibinfo {author} {\bibfnamefont {M.}~\bibnamefont
  {Leibscher}}, \bibinfo {author} {\bibfnamefont {I.~S.}\ \bibnamefont
  {Averbukh}}, \bibinfo {author} {\bibfnamefont {P.}~\bibnamefont {Rozmej}}, \
  and\ \bibinfo {author} {\bibfnamefont {R.}~\bibnamefont {Arvieu}},\ }\href
  {\doibase 10.1103/PhysRevA.69.032102} {\bibfield  {journal} {\bibinfo
  {journal} {Phys. Rev. A}\ }\textbf {\bibinfo {volume} {69}},\ \bibinfo
  {pages} {032102} (\bibinfo {year} {2004})}\BibitemShut {NoStop}%
\bibitem [{\citenamefont {Daems}\ \emph
  {et~al.}(2005{\natexlab{a}})\citenamefont {Daems}, \citenamefont
  {Gu{\'{e}}rin}, \citenamefont {Sugny},\ and\ \citenamefont
  {Jauslin}}]{Daems2005}%
  \BibitemOpen
  \bibfield  {author} {\bibinfo {author} {\bibfnamefont {D.}~\bibnamefont
  {Daems}}, \bibinfo {author} {\bibfnamefont {S.}~\bibnamefont {Gu{\'{e}}rin}},
  \bibinfo {author} {\bibfnamefont {D.}~\bibnamefont {Sugny}}, \ and\ \bibinfo
  {author} {\bibfnamefont {H.~R.}\ \bibnamefont {Jauslin}},\ }\href {\doibase
  10.1103/PhysRevLett.94.153003} {\bibfield  {journal} {\bibinfo  {journal}
  {Physical Review Letters}\ }\textbf {\bibinfo {volume} {94}},\ \bibinfo
  {pages} {153003} (\bibinfo {year} {2005}{\natexlab{a}})}\BibitemShut
  {NoStop}%
\bibitem [{\citenamefont {Torres}\ \emph {et~al.}(2007)\citenamefont {Torres},
  \citenamefont {Kajumba}, \citenamefont {Underwood}, \citenamefont {Robinson},
  \citenamefont {Baker}, \citenamefont {Tisch}, \citenamefont {de~Nalda},
  \citenamefont {Bryan}, \citenamefont {Velotta}, \citenamefont {Altucci},
  \citenamefont {Turcu},\ and\ \citenamefont {Marangos}}]{AlignmentHHG2007}%
  \BibitemOpen
  \bibfield  {author} {\bibinfo {author} {\bibfnamefont {R.}~\bibnamefont
  {Torres}}, \bibinfo {author} {\bibfnamefont {N.}~\bibnamefont {Kajumba}},
  \bibinfo {author} {\bibfnamefont {J.~G.}\ \bibnamefont {Underwood}}, \bibinfo
  {author} {\bibfnamefont {J.~S.}\ \bibnamefont {Robinson}}, \bibinfo {author}
  {\bibfnamefont {S.}~\bibnamefont {Baker}}, \bibinfo {author} {\bibfnamefont
  {J.~W.~G.}\ \bibnamefont {Tisch}}, \bibinfo {author} {\bibfnamefont
  {R.}~\bibnamefont {de~Nalda}}, \bibinfo {author} {\bibfnamefont {W.~A.}\
  \bibnamefont {Bryan}}, \bibinfo {author} {\bibfnamefont {R.}~\bibnamefont
  {Velotta}}, \bibinfo {author} {\bibfnamefont {C.}~\bibnamefont {Altucci}},
  \bibinfo {author} {\bibfnamefont {I.~C.~E.}\ \bibnamefont {Turcu}}, \ and\
  \bibinfo {author} {\bibfnamefont {J.~P.}\ \bibnamefont {Marangos}},\ }\href
  {\doibase 10.1103/PhysRevLett.98.203007} {\bibfield  {journal} {\bibinfo
  {journal} {Phys. Rev. Lett.}\ }\textbf {\bibinfo {volume} {98}},\ \bibinfo
  {pages} {203007} (\bibinfo {year} {2007})}\BibitemShut {NoStop}%
\bibitem [{\citenamefont {Ospelkaus}\ \emph {et~al.}(2010)\citenamefont
  {Ospelkaus}, \citenamefont {Ni}, \citenamefont {Qu\'em\'ener}, \citenamefont
  {Neyenhuis}, \citenamefont {Wang}, \citenamefont {de~Miranda}, \citenamefont
  {Bohn}, \citenamefont {Ye},\ and\ \citenamefont {Jin}}]{RovibronicHyperfine}%
  \BibitemOpen
  \bibfield  {author} {\bibinfo {author} {\bibfnamefont {S.}~\bibnamefont
  {Ospelkaus}}, \bibinfo {author} {\bibfnamefont {K.-K.}\ \bibnamefont {Ni}},
  \bibinfo {author} {\bibfnamefont {G.}~\bibnamefont {Qu\'em\'ener}}, \bibinfo
  {author} {\bibfnamefont {B.}~\bibnamefont {Neyenhuis}}, \bibinfo {author}
  {\bibfnamefont {D.}~\bibnamefont {Wang}}, \bibinfo {author} {\bibfnamefont
  {M.~H.~G.}\ \bibnamefont {de~Miranda}}, \bibinfo {author} {\bibfnamefont
  {J.~L.}\ \bibnamefont {Bohn}}, \bibinfo {author} {\bibfnamefont
  {J.}~\bibnamefont {Ye}}, \ and\ \bibinfo {author} {\bibfnamefont {D.~S.}\
  \bibnamefont {Jin}},\ }\href {\doibase 10.1103/PhysRevLett.104.030402}
  {\bibfield  {journal} {\bibinfo  {journal} {Phys. Rev. Lett.}\ }\textbf
  {\bibinfo {volume} {104}},\ \bibinfo {pages} {030402} (\bibinfo {year}
  {2010})}\BibitemShut {NoStop}%
\bibitem [{\citenamefont {DeMille}(2002)}]{Quantum_Info0}%
  \BibitemOpen
  \bibfield  {author} {\bibinfo {author} {\bibfnamefont {D.}~\bibnamefont
  {DeMille}},\ }\href {\doibase 10.1103/PhysRevLett.88.067901} {\bibfield
  {journal} {\bibinfo  {journal} {Phys. Rev. Lett.}\ }\textbf {\bibinfo
  {volume} {88}},\ \bibinfo {pages} {067901} (\bibinfo {year}
  {2002})}\BibitemShut {NoStop}%
\bibitem [{\citenamefont {Shapiro}\ \emph {et~al.}(2003)\citenamefont
  {Shapiro}, \citenamefont {Khavkine}, \citenamefont {Spanner},\ and\
  \citenamefont {Ivanov}}]{Quantum_Info3}%
  \BibitemOpen
  \bibfield  {author} {\bibinfo {author} {\bibfnamefont {E.~A.}\ \bibnamefont
  {Shapiro}}, \bibinfo {author} {\bibfnamefont {I.}~\bibnamefont {Khavkine}},
  \bibinfo {author} {\bibfnamefont {M.}~\bibnamefont {Spanner}}, \ and\
  \bibinfo {author} {\bibfnamefont {M.~Y.}\ \bibnamefont {Ivanov}},\ }\href
  {\doibase 10.1103/PhysRevA.67.013406} {\bibfield  {journal} {\bibinfo
  {journal} {Phys. Rev. A}\ }\textbf {\bibinfo {volume} {67}},\ \bibinfo
  {pages} {013406} (\bibinfo {year} {2003})}\BibitemShut {NoStop}%
\bibitem [{\citenamefont {Kotochigova}\ and\ \citenamefont
  {Tiesinga}(2006)}]{Quantum_Info2}%
  \BibitemOpen
  \bibfield  {author} {\bibinfo {author} {\bibfnamefont {S.}~\bibnamefont
  {Kotochigova}}\ and\ \bibinfo {author} {\bibfnamefont {E.}~\bibnamefont
  {Tiesinga}},\ }\href {\doibase 10.1103/PhysRevA.73.041405} {\bibfield
  {journal} {\bibinfo  {journal} {Phys. Rev. A}\ }\textbf {\bibinfo {volume}
  {73}},\ \bibinfo {pages} {041405} (\bibinfo {year} {2006})}\BibitemShut
  {NoStop}%
\bibitem [{\citenamefont {Arai}\ and\ \citenamefont
  {Ohtsuki}(2015)}]{Quantum_Info1}%
  \BibitemOpen
  \bibfield  {author} {\bibinfo {author} {\bibfnamefont {K.}~\bibnamefont
  {Arai}}\ and\ \bibinfo {author} {\bibfnamefont {Y.}~\bibnamefont {Ohtsuki}},\
  }\href {\doibase 10.1103/PhysRevA.92.062302} {\bibfield  {journal} {\bibinfo
  {journal} {Phys. Rev. A}\ }\textbf {\bibinfo {volume} {92}},\ \bibinfo
  {pages} {062302} (\bibinfo {year} {2015})}\BibitemShut {NoStop}%
\bibitem [{\citenamefont {Karra}\ \emph {et~al.}(2016)\citenamefont {Karra},
  \citenamefont {Sharma}, \citenamefont {Friedrich}, \citenamefont {Kais},\
  and\ \citenamefont {Herschbach}}]{Quantum_Info4}%
  \BibitemOpen
  \bibfield  {author} {\bibinfo {author} {\bibfnamefont {M.}~\bibnamefont
  {Karra}}, \bibinfo {author} {\bibfnamefont {K.}~\bibnamefont {Sharma}},
  \bibinfo {author} {\bibfnamefont {B.}~\bibnamefont {Friedrich}}, \bibinfo
  {author} {\bibfnamefont {S.}~\bibnamefont {Kais}}, \ and\ \bibinfo {author}
  {\bibfnamefont {D.}~\bibnamefont {Herschbach}},\ }\href {\doibase
  10.1063/1.4942928} {\bibfield  {journal} {\bibinfo  {journal} {The Journal of
  Chemical Physics}\ }\textbf {\bibinfo {volume} {144}},\ \bibinfo {pages}
  {094301} (\bibinfo {year} {2016})},\ \Eprint
  {http://arxiv.org/abs/https://doi.org/10.1063/1.4942928}
  {https://doi.org/10.1063/1.4942928} \BibitemShut {NoStop}%
\bibitem [{\citenamefont {Flo\ss{}}\ and\ \citenamefont
  {Averbukh}(2014)}]{Floss_Averbukh1}%
  \BibitemOpen
  \bibfield  {author} {\bibinfo {author} {\bibfnamefont {J.}~\bibnamefont
  {Flo\ss{}}}\ and\ \bibinfo {author} {\bibfnamefont {I.~S.}\ \bibnamefont
  {Averbukh}},\ }\href {\doibase 10.1103/PhysRevLett.113.043002} {\bibfield
  {journal} {\bibinfo  {journal} {Phys. Rev. Lett.}\ }\textbf {\bibinfo
  {volume} {113}},\ \bibinfo {pages} {043002} (\bibinfo {year}
  {2014})}\BibitemShut {NoStop}%
\bibitem [{\citenamefont {Flo\ss{}}\ \emph {et~al.}(2015)\citenamefont
  {Flo\ss{}}, \citenamefont {Kamalov}, \citenamefont {Averbukh},\ and\
  \citenamefont {Bucksbaum}}]{Floss_Averbukh2}%
  \BibitemOpen
  \bibfield  {author} {\bibinfo {author} {\bibfnamefont {J.}~\bibnamefont
  {Flo\ss{}}}, \bibinfo {author} {\bibfnamefont {A.}~\bibnamefont {Kamalov}},
  \bibinfo {author} {\bibfnamefont {I.~S.}\ \bibnamefont {Averbukh}}, \ and\
  \bibinfo {author} {\bibfnamefont {P.~H.}\ \bibnamefont {Bucksbaum}},\ }\href
  {\doibase 10.1103/PhysRevLett.115.203002} {\bibfield  {journal} {\bibinfo
  {journal} {Phys. Rev. Lett.}\ }\textbf {\bibinfo {volume} {115}},\ \bibinfo
  {pages} {203002} (\bibinfo {year} {2015})}\BibitemShut {NoStop}%
\bibitem [{\citenamefont {Bitter}\ and\ \citenamefont
  {Milner}(2016)}]{Bitter_Milner1}%
  \BibitemOpen
  \bibfield  {author} {\bibinfo {author} {\bibfnamefont {M.}~\bibnamefont
  {Bitter}}\ and\ \bibinfo {author} {\bibfnamefont {V.}~\bibnamefont
  {Milner}},\ }\href {\doibase 10.1103/PhysRevLett.117.144104} {\bibfield
  {journal} {\bibinfo  {journal} {Phys. Rev. Lett.}\ }\textbf {\bibinfo
  {volume} {117}},\ \bibinfo {pages} {144104} (\bibinfo {year}
  {2016})}\BibitemShut {NoStop}%
\bibitem [{\citenamefont {Bitter}\ and\ \citenamefont
  {Milner}(2017)}]{Bitter_Milner2}%
  \BibitemOpen
  \bibfield  {author} {\bibinfo {author} {\bibfnamefont {M.}~\bibnamefont
  {Bitter}}\ and\ \bibinfo {author} {\bibfnamefont {V.}~\bibnamefont
  {Milner}},\ }\href {\doibase 10.1103/PhysRevLett.118.034101} {\bibfield
  {journal} {\bibinfo  {journal} {Phys. Rev. Lett.}\ }\textbf {\bibinfo
  {volume} {118}},\ \bibinfo {pages} {034101} (\bibinfo {year}
  {2017})}\BibitemShut {NoStop}%
\bibitem [{\citenamefont {Zhdanovich}\ \emph {et~al.}(2011)\citenamefont
  {Zhdanovich}, \citenamefont {Milner}, \citenamefont {Bloomquist},
  \citenamefont {Flo\ss{}}, \citenamefont {Averbukh}, \citenamefont {Hepburn},\
  and\ \citenamefont {Milner}}]{UltrashortQCMilner2011}%
  \BibitemOpen
  \bibfield  {author} {\bibinfo {author} {\bibfnamefont {S.}~\bibnamefont
  {Zhdanovich}}, \bibinfo {author} {\bibfnamefont {A.~A.}\ \bibnamefont
  {Milner}}, \bibinfo {author} {\bibfnamefont {C.}~\bibnamefont {Bloomquist}},
  \bibinfo {author} {\bibfnamefont {J.}~\bibnamefont {Flo\ss{}}}, \bibinfo
  {author} {\bibfnamefont {I.~S.}\ \bibnamefont {Averbukh}}, \bibinfo {author}
  {\bibfnamefont {J.~W.}\ \bibnamefont {Hepburn}}, \ and\ \bibinfo {author}
  {\bibfnamefont {V.}~\bibnamefont {Milner}},\ }\href {\doibase
  10.1103/PhysRevLett.107.243004} {\bibfield  {journal} {\bibinfo  {journal}
  {Phys. Rev. Lett.}\ }\textbf {\bibinfo {volume} {107}},\ \bibinfo {pages}
  {243004} (\bibinfo {year} {2011})}\BibitemShut {NoStop}%
\bibitem [{\citenamefont {Friedrich}\ and\ \citenamefont
  {Herschbach}(1991)}]{PendularNature}%
  \BibitemOpen
  \bibfield  {author} {\bibinfo {author} {\bibfnamefont {B.}~\bibnamefont
  {Friedrich}}\ and\ \bibinfo {author} {\bibfnamefont {D.~R.}\ \bibnamefont
  {Herschbach}},\ }\href {\doibase 10.1038/353412a0} {\bibfield  {journal}
  {\bibinfo  {journal} {Nature}\ }\textbf {\bibinfo {volume} {353}},\ \bibinfo
  {pages} {412} (\bibinfo {year} {1991})}\BibitemShut {NoStop}%
\bibitem [{\citenamefont {Rost}\ \emph {et~al.}(1992)\citenamefont {Rost},
  \citenamefont {Griffin}, \citenamefont {Friedrich},\ and\ \citenamefont
  {Herschbach}}]{PendularStatesPRL}%
  \BibitemOpen
  \bibfield  {author} {\bibinfo {author} {\bibfnamefont {J.~M.}\ \bibnamefont
  {Rost}}, \bibinfo {author} {\bibfnamefont {J.~C.}\ \bibnamefont {Griffin}},
  \bibinfo {author} {\bibfnamefont {B.}~\bibnamefont {Friedrich}}, \ and\
  \bibinfo {author} {\bibfnamefont {D.~R.}\ \bibnamefont {Herschbach}},\ }\href
  {\doibase 10.1103/PhysRevLett.68.1299} {\bibfield  {journal} {\bibinfo
  {journal} {Phys. Rev. Lett.}\ }\textbf {\bibinfo {volume} {68}},\ \bibinfo
  {pages} {1299} (\bibinfo {year} {1992})}\BibitemShut {NoStop}%
\bibitem [{\citenamefont {Auzinsh}\ and\ \citenamefont
  {Ferber}(1992)}]{AuzinshStark1992}%
  \BibitemOpen
  \bibfield  {author} {\bibinfo {author} {\bibfnamefont {M.~P.}\ \bibnamefont
  {Auzinsh}}\ and\ \bibinfo {author} {\bibfnamefont {R.~S.}\ \bibnamefont
  {Ferber}},\ }\href {\doibase 10.1103/PhysRevLett.69.3463} {\bibfield
  {journal} {\bibinfo  {journal} {Phys. Rev. Lett.}\ }\textbf {\bibinfo
  {volume} {69}},\ \bibinfo {pages} {3463} (\bibinfo {year}
  {1992})}\BibitemShut {NoStop}%
\bibitem [{\citenamefont {Friedrich}\ and\ \citenamefont
  {Herschbach}(1999{\natexlab{a}})}]{Friedrich1999}%
  \BibitemOpen
  \bibfield  {author} {\bibinfo {author} {\bibfnamefont {B.}~\bibnamefont
  {Friedrich}}\ and\ \bibinfo {author} {\bibfnamefont {D.}~\bibnamefont
  {Herschbach}},\ }\href {\doibase 10.1063/1.479917} {\bibfield  {journal}
  {\bibinfo  {journal} {J. Chem. Phys.}\ }\textbf {\bibinfo {volume} {111}},\
  \bibinfo {pages} {6157} (\bibinfo {year} {1999}{\natexlab{a}})}\BibitemShut
  {NoStop}%
\bibitem [{\citenamefont {Friedrich}\ and\ \citenamefont
  {Herschbach}(1999{\natexlab{b}})}]{Friedrich1999a}%
  \BibitemOpen
  \bibfield  {author} {\bibinfo {author} {\bibfnamefont {B.}~\bibnamefont
  {Friedrich}}\ and\ \bibinfo {author} {\bibnamefont {Herschbach}},\ }\href
  {\doibase 10.1021/jp992131w} {\bibfield  {journal} {\bibinfo  {journal} {J.
  Phys. Chem. A}\ }\textbf {\bibinfo {volume} {103}},\ \bibinfo {pages} {10280}
  (\bibinfo {year} {1999}{\natexlab{b}})}\BibitemShut {NoStop}%
\bibitem [{\citenamefont {Schmidt}\ and\ \citenamefont
  {Friedrich}(2014{\natexlab{a}})}]{TI4}%
  \BibitemOpen
  \bibfield  {author} {\bibinfo {author} {\bibfnamefont {B.}~\bibnamefont
  {Schmidt}}\ and\ \bibinfo {author} {\bibfnamefont {B.}~\bibnamefont
  {Friedrich}},\ }\href {\doibase 10.1063/1.4864465} {\bibfield  {journal}
  {\bibinfo  {journal} {The Journal of Chemical Physics}\ }\textbf {\bibinfo
  {volume} {140}},\ \bibinfo {pages} {064317} (\bibinfo {year}
  {2014}{\natexlab{a}})},\ \Eprint
  {http://arxiv.org/abs/https://doi.org/10.1063/1.4864465}
  {https://doi.org/10.1063/1.4864465} \BibitemShut {NoStop}%
\bibitem [{\citenamefont {Schmidt}\ and\ \citenamefont
  {Friedrich}(2015)}]{TI3}%
  \BibitemOpen
  \bibfield  {author} {\bibinfo {author} {\bibfnamefont {B.}~\bibnamefont
  {Schmidt}}\ and\ \bibinfo {author} {\bibfnamefont {B.}~\bibnamefont
  {Friedrich}},\ }\href {\doibase 10.1103/PhysRevA.91.022111} {\bibfield
  {journal} {\bibinfo  {journal} {Phys. Rev. A}\ }\textbf {\bibinfo {volume}
  {91}},\ \bibinfo {pages} {022111} (\bibinfo {year} {2015})}\BibitemShut
  {NoStop}%
\bibitem [{\citenamefont {Becker}\ \emph {et~al.}(2017)\citenamefont {Becker},
  \citenamefont {Mirahmadi}, \citenamefont {Schmidt}, \citenamefont {Schatz},\
  and\ \citenamefont {Friedrich}}]{TI2}%
  \BibitemOpen
  \bibfield  {author} {\bibinfo {author} {\bibfnamefont {S.}~\bibnamefont
  {Becker}}, \bibinfo {author} {\bibfnamefont {M.}~\bibnamefont {Mirahmadi}},
  \bibinfo {author} {\bibfnamefont {B.}~\bibnamefont {Schmidt}}, \bibinfo
  {author} {\bibfnamefont {K.}~\bibnamefont {Schatz}}, \ and\ \bibinfo {author}
  {\bibfnamefont {B.}~\bibnamefont {Friedrich}},\ }\href {\doibase
  10.1140/epjd/e2017-80134-6} {\bibfield  {journal} {\bibinfo  {journal} {The
  European Physical Journal D}\ }\textbf {\bibinfo {volume} {71}},\ \bibinfo
  {pages} {149} (\bibinfo {year} {2017})}\BibitemShut {NoStop}%
\bibitem [{\citenamefont {Schatz}\ \emph {et~al.}(2018)\citenamefont {Schatz},
  \citenamefont {Friedrich}, \citenamefont {Becker},\ and\ \citenamefont
  {Schmidt}}]{TI1}%
  \BibitemOpen
  \bibfield  {author} {\bibinfo {author} {\bibfnamefont {K.}~\bibnamefont
  {Schatz}}, \bibinfo {author} {\bibfnamefont {B.}~\bibnamefont {Friedrich}},
  \bibinfo {author} {\bibfnamefont {S.}~\bibnamefont {Becker}}, \ and\ \bibinfo
  {author} {\bibfnamefont {B.}~\bibnamefont {Schmidt}},\ }\href {\doibase
  10.1103/PhysRevA.97.053417} {\bibfield  {journal} {\bibinfo  {journal} {Phys.
  Rev. A}\ }\textbf {\bibinfo {volume} {97}},\ \bibinfo {pages} {053417}
  (\bibinfo {year} {2018})}\BibitemShut {NoStop}%
\bibitem [{\citenamefont {Schmidt}\ and\ \citenamefont
  {Friedrich}(2014{\natexlab{b}})}]{TI0}%
  \BibitemOpen
  \bibfield  {author} {\bibinfo {author} {\bibfnamefont {B.}~\bibnamefont
  {Schmidt}}\ and\ \bibinfo {author} {\bibfnamefont {B.}~\bibnamefont
  {Friedrich}},\ }\href {\doibase 10.3389/fphy.2014.00037} {\bibfield
  {journal} {\bibinfo  {journal} {Frontiers in Physics}\ }\textbf {\bibinfo
  {volume} {2}},\ \bibinfo {pages} {37} (\bibinfo {year}
  {2014}{\natexlab{b}})}\BibitemShut {NoStop}%
\bibitem [{\citenamefont {Li}\ \emph {et~al.}(2017)\citenamefont {Li},
  \citenamefont {Petrov}, \citenamefont {Makrides}, \citenamefont {Tiesinga},\
  and\ \citenamefont {Kotochigova}}]{DC_Manipulation2017}%
  \BibitemOpen
  \bibfield  {author} {\bibinfo {author} {\bibfnamefont {M.}~\bibnamefont
  {Li}}, \bibinfo {author} {\bibfnamefont {A.}~\bibnamefont {Petrov}}, \bibinfo
  {author} {\bibfnamefont {C.}~\bibnamefont {Makrides}}, \bibinfo {author}
  {\bibfnamefont {E.}~\bibnamefont {Tiesinga}}, \ and\ \bibinfo {author}
  {\bibfnamefont {S.}~\bibnamefont {Kotochigova}},\ }\href {\doibase
  10.1103/PhysRevA.95.063422} {\bibfield  {journal} {\bibinfo  {journal} {Phys.
  Rev. A}\ }\textbf {\bibinfo {volume} {95}},\ \bibinfo {pages} {063422}
  (\bibinfo {year} {2017})}\BibitemShut {NoStop}%
\bibitem [{\citenamefont {Kang}\ \emph {et~al.}(2018)\citenamefont {Kang},
  \citenamefont {Park}, \citenamefont {Kim},\ and\ \citenamefont
  {Kang}}]{DCManipulation2018}%
  \BibitemOpen
  \bibfield  {author} {\bibinfo {author} {\bibfnamefont {H.}~\bibnamefont
  {Kang}}, \bibinfo {author} {\bibfnamefont {Y.}~\bibnamefont {Park}}, \bibinfo
  {author} {\bibfnamefont {Z.~H.}\ \bibnamefont {Kim}}, \ and\ \bibinfo
  {author} {\bibfnamefont {H.}~\bibnamefont {Kang}},\ }\href {\doibase
  10.1021/acs.jpca.7b11740} {\bibfield  {journal} {\bibinfo  {journal} {The
  Journal of Physical Chemistry A}\ }\textbf {\bibinfo {volume} {122}},\
  \bibinfo {pages} {2871} (\bibinfo {year} {2018})}\BibitemShut {NoStop}%
\bibitem [{\citenamefont {Park}\ \emph {et~al.}(2021)\citenamefont {Park},
  \citenamefont {Shin},\ and\ \citenamefont
  {Kang}}]{DCManipulation_Account2021}%
  \BibitemOpen
  \bibfield  {author} {\bibinfo {author} {\bibfnamefont {Y.}~\bibnamefont
  {Park}}, \bibinfo {author} {\bibfnamefont {S.}~\bibnamefont {Shin}}, \ and\
  \bibinfo {author} {\bibfnamefont {H.}~\bibnamefont {Kang}},\ }\href {\doibase
  10.1021/acs.accounts.0c00609} {\bibfield  {journal} {\bibinfo  {journal}
  {Accounts of Chemical Research}\ }\textbf {\bibinfo {volume} {54}},\ \bibinfo
  {pages} {323} (\bibinfo {year} {2021})}\BibitemShut {NoStop}%
\bibitem [{\citenamefont {Stapelfeldt}\ and\ \citenamefont
  {Seideman}(2003)}]{RevSeidemannStapelfeldt}%
  \BibitemOpen
  \bibfield  {author} {\bibinfo {author} {\bibfnamefont {H.}~\bibnamefont
  {Stapelfeldt}}\ and\ \bibinfo {author} {\bibfnamefont {T.}~\bibnamefont
  {Seideman}},\ }\href {\doibase 10.1103/RevModPhys.75.543} {\bibfield
  {journal} {\bibinfo  {journal} {Rev. Mod. Phys.}\ }\textbf {\bibinfo {volume}
  {75}},\ \bibinfo {pages} {543} (\bibinfo {year} {2003})}\BibitemShut
  {NoStop}%
\bibitem [{\citenamefont {Seideman}(1999)}]{Seideman_revival}%
  \BibitemOpen
  \bibfield  {author} {\bibinfo {author} {\bibfnamefont {T.}~\bibnamefont
  {Seideman}},\ }\href {\doibase 10.1103/PhysRevLett.83.4971} {\bibfield
  {journal} {\bibinfo  {journal} {Phys. Rev. Lett.}\ }\textbf {\bibinfo
  {volume} {83}},\ \bibinfo {pages} {4971} (\bibinfo {year}
  {1999})}\BibitemShut {NoStop}%
\bibitem [{\citenamefont {Cai}\ and\ \citenamefont
  {Friedrich}(2001)}]{Cai2001}%
  \BibitemOpen
  \bibfield  {author} {\bibinfo {author} {\bibfnamefont {L.}~\bibnamefont
  {Cai}}\ and\ \bibinfo {author} {\bibfnamefont {B.}~\bibnamefont
  {Friedrich}},\ }\href {\doibase 10.1135/cccc20010991} {\bibfield  {journal}
  {\bibinfo  {journal} {Collect. Czechoslov. Chem. Commun.}\ }\textbf {\bibinfo
  {volume} {66}},\ \bibinfo {pages} {991} (\bibinfo {year} {2001})}\BibitemShut
  {NoStop}%
\bibitem [{\citenamefont {Cai}\ \emph {et~al.}(2001)\citenamefont {Cai},
  \citenamefont {Marango},\ and\ \citenamefont {Friedrich}}]{Cai_BF_2001}%
  \BibitemOpen
  \bibfield  {author} {\bibinfo {author} {\bibfnamefont {L.}~\bibnamefont
  {Cai}}, \bibinfo {author} {\bibfnamefont {J.}~\bibnamefont {Marango}}, \ and\
  \bibinfo {author} {\bibfnamefont {B.}~\bibnamefont {Friedrich}},\ }\href
  {\doibase 10.1103/PhysRevLett.86.775} {\bibfield  {journal} {\bibinfo
  {journal} {Phys. Rev. Lett.}\ }\textbf {\bibinfo {volume} {86}},\ \bibinfo
  {pages} {775} (\bibinfo {year} {2001})}\BibitemShut {NoStop}%
\bibitem [{\citenamefont {Sugny}\ \emph {et~al.}(2005)\citenamefont {Sugny},
  \citenamefont {Keller}, \citenamefont {Atabek}, \citenamefont {Daems},
  \citenamefont {Dion}, \citenamefont {Gu\'erin},\ and\ \citenamefont
  {Jauslin}}]{Jauslin2005}%
  \BibitemOpen
  \bibfield  {author} {\bibinfo {author} {\bibfnamefont {D.}~\bibnamefont
  {Sugny}}, \bibinfo {author} {\bibfnamefont {A.}~\bibnamefont {Keller}},
  \bibinfo {author} {\bibfnamefont {O.}~\bibnamefont {Atabek}}, \bibinfo
  {author} {\bibfnamefont {D.}~\bibnamefont {Daems}}, \bibinfo {author}
  {\bibfnamefont {C.~M.}\ \bibnamefont {Dion}}, \bibinfo {author}
  {\bibfnamefont {S.}~\bibnamefont {Gu\'erin}}, \ and\ \bibinfo {author}
  {\bibfnamefont {H.~R.}\ \bibnamefont {Jauslin}},\ }\href {\doibase
  10.1103/PhysRevA.71.063402} {\bibfield  {journal} {\bibinfo  {journal} {Phys.
  Rev. A}\ }\textbf {\bibinfo {volume} {71}},\ \bibinfo {pages} {063402}
  (\bibinfo {year} {2005})}\BibitemShut {NoStop}%
\bibitem [{\citenamefont {Daems}\ \emph
  {et~al.}(2005{\natexlab{b}})\citenamefont {Daems}, \citenamefont {Gu\'erin},
  \citenamefont {Hertz}, \citenamefont {Jauslin}, \citenamefont {Lavorel},\
  and\ \citenamefont {Faucher}}]{EllipticLaserPulses2005}%
  \BibitemOpen
  \bibfield  {author} {\bibinfo {author} {\bibfnamefont {D.}~\bibnamefont
  {Daems}}, \bibinfo {author} {\bibfnamefont {S.}~\bibnamefont {Gu\'erin}},
  \bibinfo {author} {\bibfnamefont {E.}~\bibnamefont {Hertz}}, \bibinfo
  {author} {\bibfnamefont {H.~R.}\ \bibnamefont {Jauslin}}, \bibinfo {author}
  {\bibfnamefont {B.}~\bibnamefont {Lavorel}}, \ and\ \bibinfo {author}
  {\bibfnamefont {O.}~\bibnamefont {Faucher}},\ }\href {\doibase
  10.1103/PhysRevLett.95.063005} {\bibfield  {journal} {\bibinfo  {journal}
  {Phys. Rev. Lett.}\ }\textbf {\bibinfo {volume} {95}},\ \bibinfo {pages}
  {063005} (\bibinfo {year} {2005}{\natexlab{b}})}\BibitemShut {NoStop}%
\bibitem [{\citenamefont {Hertz}\ \emph
  {et~al.}(2007{\natexlab{a}})\citenamefont {Hertz}, \citenamefont {Rouz\'ee},
  \citenamefont {Gu\'erin}, \citenamefont {Lavorel},\ and\ \citenamefont
  {Faucher}}]{Rouzee2007}%
  \BibitemOpen
  \bibfield  {author} {\bibinfo {author} {\bibfnamefont {E.}~\bibnamefont
  {Hertz}}, \bibinfo {author} {\bibfnamefont {A.}~\bibnamefont {Rouz\'ee}},
  \bibinfo {author} {\bibfnamefont {S.}~\bibnamefont {Gu\'erin}}, \bibinfo
  {author} {\bibfnamefont {B.}~\bibnamefont {Lavorel}}, \ and\ \bibinfo
  {author} {\bibfnamefont {O.}~\bibnamefont {Faucher}},\ }\href {\doibase
  10.1103/PhysRevA.75.031403} {\bibfield  {journal} {\bibinfo  {journal} {Phys.
  Rev. A}\ }\textbf {\bibinfo {volume} {75}},\ \bibinfo {pages} {031403}
  (\bibinfo {year} {2007}{\natexlab{a}})}\BibitemShut {NoStop}%
\bibitem [{\citenamefont {Hertz}\ \emph
  {et~al.}(2007{\natexlab{b}})\citenamefont {Hertz}, \citenamefont {Daems},
  \citenamefont {Gu\'erin}, \citenamefont {Jauslin}, \citenamefont {Lavorel},\
  and\ \citenamefont {Faucher}}]{FieldFreeDaemsGuerin2007}%
  \BibitemOpen
  \bibfield  {author} {\bibinfo {author} {\bibfnamefont {E.}~\bibnamefont
  {Hertz}}, \bibinfo {author} {\bibfnamefont {D.}~\bibnamefont {Daems}},
  \bibinfo {author} {\bibfnamefont {S.}~\bibnamefont {Gu\'erin}}, \bibinfo
  {author} {\bibfnamefont {H.~R.}\ \bibnamefont {Jauslin}}, \bibinfo {author}
  {\bibfnamefont {B.}~\bibnamefont {Lavorel}}, \ and\ \bibinfo {author}
  {\bibfnamefont {O.}~\bibnamefont {Faucher}},\ }\href {\doibase
  10.1103/PhysRevA.76.043423} {\bibfield  {journal} {\bibinfo  {journal} {Phys.
  Rev. A}\ }\textbf {\bibinfo {volume} {76}},\ \bibinfo {pages} {043423}
  (\bibinfo {year} {2007}{\natexlab{b}})}\BibitemShut {NoStop}%
\bibitem [{\citenamefont {Gu\'erin}\ \emph {et~al.}(2008)\citenamefont
  {Gu\'erin}, \citenamefont {Rouz\'ee},\ and\ \citenamefont
  {Hertz}}]{GuerinRouzee2008}%
  \BibitemOpen
  \bibfield  {author} {\bibinfo {author} {\bibfnamefont {S.}~\bibnamefont
  {Gu\'erin}}, \bibinfo {author} {\bibfnamefont {A.}~\bibnamefont {Rouz\'ee}},
  \ and\ \bibinfo {author} {\bibfnamefont {E.}~\bibnamefont {Hertz}},\ }\href
  {\doibase 10.1103/PhysRevA.77.041404} {\bibfield  {journal} {\bibinfo
  {journal} {Phys. Rev. A}\ }\textbf {\bibinfo {volume} {77}},\ \bibinfo
  {pages} {041404} (\bibinfo {year} {2008})}\BibitemShut {NoStop}%
\bibitem [{\citenamefont {Owschimikow}\ \emph {et~al.}(2009)\citenamefont
  {Owschimikow}, \citenamefont {Schmidt},\ and\ \citenamefont
  {Schwentner}}]{Owschimikow2009}%
  \BibitemOpen
  \bibfield  {author} {\bibinfo {author} {\bibfnamefont {N.}~\bibnamefont
  {Owschimikow}}, \bibinfo {author} {\bibfnamefont {B.}~\bibnamefont
  {Schmidt}}, \ and\ \bibinfo {author} {\bibfnamefont {N.}~\bibnamefont
  {Schwentner}},\ }\href {\doibase 10.1103/PhysRevA.80.053409} {\bibfield
  {journal} {\bibinfo  {journal} {Phys. Rev. A}\ }\textbf {\bibinfo {volume}
  {80}},\ \bibinfo {pages} {053409} (\bibinfo {year} {2009})}\BibitemShut
  {NoStop}%
\bibitem [{\citenamefont {Owschimikow}\ \emph {et~al.}(2011)\citenamefont
  {Owschimikow}, \citenamefont {Schmidt},\ and\ \citenamefont
  {Schwentner}}]{Owschimikow2011}%
  \BibitemOpen
  \bibfield  {author} {\bibinfo {author} {\bibfnamefont {N.}~\bibnamefont
  {Owschimikow}}, \bibinfo {author} {\bibfnamefont {B.}~\bibnamefont
  {Schmidt}}, \ and\ \bibinfo {author} {\bibfnamefont {N.}~\bibnamefont
  {Schwentner}},\ }\href {\doibase 10.1039/c0cp02260h} {\bibfield  {journal}
  {\bibinfo  {journal} {Phys. Chem. Chem. Phys.}\ }\textbf {\bibinfo {volume}
  {13}},\ \bibinfo {pages} {8671} (\bibinfo {year} {2011})}\BibitemShut
  {NoStop}%
\bibitem [{\citenamefont {Fleischer}\ \emph {et~al.}(2012)\citenamefont
  {Fleischer}, \citenamefont {Khodorkovsky}, \citenamefont {Gershnabel},
  \citenamefont {Prior},\ and\ \citenamefont
  {Averbukh}}]{FleischerAverbukh2012}%
  \BibitemOpen
  \bibfield  {author} {\bibinfo {author} {\bibfnamefont {S.}~\bibnamefont
  {Fleischer}}, \bibinfo {author} {\bibfnamefont {Y.}~\bibnamefont
  {Khodorkovsky}}, \bibinfo {author} {\bibfnamefont {E.}~\bibnamefont
  {Gershnabel}}, \bibinfo {author} {\bibfnamefont {Y.}~\bibnamefont {Prior}}, \
  and\ \bibinfo {author} {\bibfnamefont {I.~S.}\ \bibnamefont {Averbukh}},\
  }\href {\doibase https://doi.org/10.1002/ijch.201100161} {\bibfield
  {journal} {\bibinfo  {journal} {Israel Journal of Chemistry}\ }\textbf
  {\bibinfo {volume} {52}},\ \bibinfo {pages} {414} (\bibinfo {year} {2012})},\
  \Eprint
  {http://arxiv.org/abs/https://onlinelibrary.wiley.com/doi/pdf/10.1002/ijch.201100161}
  {https://onlinelibrary.wiley.com/doi/pdf/10.1002/ijch.201100161} \BibitemShut
  {NoStop}%
\bibitem [{\citenamefont {Hamraoui}\ \emph {et~al.}(2017)\citenamefont
  {Hamraoui}, \citenamefont {Babilotte}, \citenamefont {Billard}, \citenamefont
  {Hertz}, \citenamefont {Faucher}, \citenamefont {Coudert}, \citenamefont
  {Sugny},\ and\ \citenamefont {Lavorel}}]{THzPulseShaping2017}%
  \BibitemOpen
  \bibfield  {author} {\bibinfo {author} {\bibfnamefont {K.}~\bibnamefont
  {Hamraoui}}, \bibinfo {author} {\bibfnamefont {P.}~\bibnamefont {Babilotte}},
  \bibinfo {author} {\bibfnamefont {F.}~\bibnamefont {Billard}}, \bibinfo
  {author} {\bibfnamefont {E.}~\bibnamefont {Hertz}}, \bibinfo {author}
  {\bibfnamefont {O.}~\bibnamefont {Faucher}}, \bibinfo {author} {\bibfnamefont
  {L.~H.}\ \bibnamefont {Coudert}}, \bibinfo {author} {\bibfnamefont
  {D.}~\bibnamefont {Sugny}}, \ and\ \bibinfo {author} {\bibfnamefont
  {B.}~\bibnamefont {Lavorel}},\ }\href {\doibase 10.1103/PhysRevA.96.043416}
  {\bibfield  {journal} {\bibinfo  {journal} {Phys. Rev. A}\ }\textbf {\bibinfo
  {volume} {96}},\ \bibinfo {pages} {043416} (\bibinfo {year}
  {2017})}\BibitemShut {NoStop}%
\bibitem [{\citenamefont {Tehini}\ \emph {et~al.}(2019)\citenamefont {Tehini},
  \citenamefont {Hamraoui},\ and\ \citenamefont {Sugny}}]{THzSugny2019}%
  \BibitemOpen
  \bibfield  {author} {\bibinfo {author} {\bibfnamefont {R.}~\bibnamefont
  {Tehini}}, \bibinfo {author} {\bibfnamefont {K.}~\bibnamefont {Hamraoui}}, \
  and\ \bibinfo {author} {\bibfnamefont {D.}~\bibnamefont {Sugny}},\ }\href
  {\doibase 10.1103/PhysRevA.99.033419} {\bibfield  {journal} {\bibinfo
  {journal} {Phys. Rev. A}\ }\textbf {\bibinfo {volume} {99}},\ \bibinfo
  {pages} {033419} (\bibinfo {year} {2019})}\BibitemShut {NoStop}%
\bibitem [{\citenamefont {Nautiyal}\ \emph {et~al.}(2021)\citenamefont
  {Nautiyal}, \citenamefont {Devi}, \citenamefont {Tyagi}, \citenamefont
  {Vidhani}, \citenamefont {Maan},\ and\ \citenamefont
  {Prasad}}]{Nautiyal2021}%
  \BibitemOpen
  \bibfield  {author} {\bibinfo {author} {\bibfnamefont {V.~V.}\ \bibnamefont
  {Nautiyal}}, \bibinfo {author} {\bibfnamefont {S.}~\bibnamefont {Devi}},
  \bibinfo {author} {\bibfnamefont {A.}~\bibnamefont {Tyagi}}, \bibinfo
  {author} {\bibfnamefont {B.}~\bibnamefont {Vidhani}}, \bibinfo {author}
  {\bibfnamefont {A.}~\bibnamefont {Maan}}, \ and\ \bibinfo {author}
  {\bibfnamefont {V.}~\bibnamefont {Prasad}},\ }\href {\doibase
  https://doi.org/10.1016/j.saa.2021.119663} {\bibfield  {journal} {\bibinfo
  {journal} {Spectrochimica Acta Part A: Molecular and Biomolecular
  Spectroscopy}\ }\textbf {\bibinfo {volume} {256}},\ \bibinfo {pages} {119663}
  (\bibinfo {year} {2021})}\BibitemShut {NoStop}%
\bibitem [{\citenamefont {Dion}\ \emph {et~al.}(2001)\citenamefont {Dion},
  \citenamefont {Keller},\ and\ \citenamefont {Atabek}}]{Dion2001}%
  \BibitemOpen
  \bibfield  {author} {\bibinfo {author} {\bibfnamefont {C.}~\bibnamefont
  {Dion}}, \bibinfo {author} {\bibfnamefont {A.}~\bibnamefont {Keller}}, \ and\
  \bibinfo {author} {\bibfnamefont {O.}~\bibnamefont {Atabek}},\ }\href
  {\doibase 10.1007/s100530170223} {\bibfield  {journal} {\bibinfo  {journal}
  {Eur. Phys. J. D}\ }\textbf {\bibinfo {volume} {14}},\ \bibinfo {pages} {249}
  (\bibinfo {year} {2001})}\BibitemShut {NoStop}%
\bibitem [{\citenamefont {Ortigoso}\ \emph {et~al.}(1999)\citenamefont
  {Ortigoso}, \citenamefont {Rodr\'{i}guez}, \citenamefont {Gupta},\ and\
  \citenamefont {Friedrich}}]{Ortigoso_BF_Dynamics}%
  \BibitemOpen
  \bibfield  {author} {\bibinfo {author} {\bibfnamefont {J.}~\bibnamefont
  {Ortigoso}}, \bibinfo {author} {\bibfnamefont {M.}~\bibnamefont
  {Rodr\'{i}guez}}, \bibinfo {author} {\bibfnamefont {M.}~\bibnamefont
  {Gupta}}, \ and\ \bibinfo {author} {\bibfnamefont {B.}~\bibnamefont
  {Friedrich}},\ }\href {\doibase 10.1063/1.478241} {\bibfield  {journal}
  {\bibinfo  {journal} {J. Chem. Phys.}\ }\textbf {\bibinfo {volume} {110}},\
  \bibinfo {pages} {3870} (\bibinfo {year} {1999})}\BibitemShut {NoStop}%
\bibitem [{\citenamefont {Mirahmadi}\ \emph {et~al.}(2018)\citenamefont
  {Mirahmadi}, \citenamefont {Schmidt}, \citenamefont {Karra},\ and\
  \citenamefont {Friedrich}}]{Mirahmadi2018}%
  \BibitemOpen
  \bibfield  {author} {\bibinfo {author} {\bibfnamefont {M.}~\bibnamefont
  {Mirahmadi}}, \bibinfo {author} {\bibfnamefont {B.}~\bibnamefont {Schmidt}},
  \bibinfo {author} {\bibfnamefont {M.}~\bibnamefont {Karra}}, \ and\ \bibinfo
  {author} {\bibfnamefont {B.}~\bibnamefont {Friedrich}},\ }\href {\doibase
  10.1063/1.5051591} {\bibfield  {journal} {\bibinfo  {journal} {The Journal of
  Chemical Physics}\ }\textbf {\bibinfo {volume} {149}},\ \bibinfo {pages}
  {174109} (\bibinfo {year} {2018})},\ \Eprint
  {http://arxiv.org/abs/https://doi.org/10.1063/1.5051591}
  {https://doi.org/10.1063/1.5051591} \BibitemShut {NoStop}%
\bibitem [{\citenamefont {Sugny}\ \emph {et~al.}(2004)\citenamefont {Sugny},
  \citenamefont {Keller}, \citenamefont {Atabek}, \citenamefont {Daems},
  \citenamefont {Gu\'erin},\ and\ \citenamefont {Jauslin}}]{TDUPT2004}%
  \BibitemOpen
  \bibfield  {author} {\bibinfo {author} {\bibfnamefont {D.}~\bibnamefont
  {Sugny}}, \bibinfo {author} {\bibfnamefont {A.}~\bibnamefont {Keller}},
  \bibinfo {author} {\bibfnamefont {O.}~\bibnamefont {Atabek}}, \bibinfo
  {author} {\bibfnamefont {D.}~\bibnamefont {Daems}}, \bibinfo {author}
  {\bibfnamefont {S.}~\bibnamefont {Gu\'erin}}, \ and\ \bibinfo {author}
  {\bibfnamefont {H.~R.}\ \bibnamefont {Jauslin}},\ }\href {\doibase
  10.1103/PhysRevA.69.043407} {\bibfield  {journal} {\bibinfo  {journal} {Phys.
  Rev. A}\ }\textbf {\bibinfo {volume} {69}},\ \bibinfo {pages} {043407}
  (\bibinfo {year} {2004})}\BibitemShut {NoStop}%
\bibitem [{\citenamefont {Dittrich}\ and\ \citenamefont
  {Reutera}(2001)}]{Superconvergent_KAM1}%
  \BibitemOpen
  \bibfield  {author} {\bibinfo {author} {\bibfnamefont {W.}~\bibnamefont
  {Dittrich}}\ and\ \bibinfo {author} {\bibfnamefont {M.}~\bibnamefont
  {Reutera}},\ }\enquote {\bibinfo {title} {{Superconvergent Perturbation
  Theory, KAM Theorem (Introduction)}},}\ in\ \href {\doibase
  10.1007/978-3-642-56430-7_14} {\emph {\bibinfo {booktitle} {Classical and
  Quantum Dynamics: From Classical Paths to Path Integrals}}}\ (\bibinfo
  {publisher} {Springer Berlin Heidelberg},\ \bibinfo {address} {Berlin,
  Heidelberg},\ \bibinfo {year} {2001})\ pp.\ \bibinfo {pages}
  {155--162}\BibitemShut {NoStop}%
\bibitem [{\citenamefont {Daems}\ \emph {et~al.}(2003)\citenamefont {Daems},
  \citenamefont {Keller}, \citenamefont {Gu\'erin}, \citenamefont {Jauslin},\
  and\ \citenamefont {Atabek}}]{Superconvergent_KAM2}%
  \BibitemOpen
  \bibfield  {author} {\bibinfo {author} {\bibfnamefont {D.}~\bibnamefont
  {Daems}}, \bibinfo {author} {\bibfnamefont {A.}~\bibnamefont {Keller}},
  \bibinfo {author} {\bibfnamefont {S.}~\bibnamefont {Gu\'erin}}, \bibinfo
  {author} {\bibfnamefont {H.~R.}\ \bibnamefont {Jauslin}}, \ and\ \bibinfo
  {author} {\bibfnamefont {O.}~\bibnamefont {Atabek}},\ }\href {\doibase
  10.1103/PhysRevA.67.052505} {\bibfield  {journal} {\bibinfo  {journal} {Phys.
  Rev. A}\ }\textbf {\bibinfo {volume} {67}},\ \bibinfo {pages} {052505}
  (\bibinfo {year} {2003})}\BibitemShut {NoStop}%
\bibitem [{\citenamefont {Schmidt}\ and\ \citenamefont
  {Lorenz}(2017)}]{Schmidt2017}%
  \BibitemOpen
  \bibfield  {author} {\bibinfo {author} {\bibfnamefont {B.}~\bibnamefont
  {Schmidt}}\ and\ \bibinfo {author} {\bibfnamefont {U.}~\bibnamefont
  {Lorenz}},\ }\href {\doibase 10.1016/j.cpc.2016.12.007} {\bibfield  {journal}
  {\bibinfo  {journal} {Comput. Phys. Commun.}\ }\textbf {\bibinfo {volume}
  {213}},\ \bibinfo {pages} {223} (\bibinfo {year} {2017})}\BibitemShut
  {NoStop}%
\bibitem [{\citenamefont {Schmidt}\ and\ \citenamefont
  {Hartmann}(2018)}]{Wavepacket2}%
  \BibitemOpen
  \bibfield  {author} {\bibinfo {author} {\bibfnamefont {B.}~\bibnamefont
  {Schmidt}}\ and\ \bibinfo {author} {\bibfnamefont {C.}~\bibnamefont
  {Hartmann}},\ }\href {\doibase https://doi.org/10.1016/j.cpc.2018.02.022}
  {\bibfield  {journal} {\bibinfo  {journal} {Computer Physics Communications}\
  }\textbf {\bibinfo {volume} {228}},\ \bibinfo {pages} {229} (\bibinfo {year}
  {2018})}\BibitemShut {NoStop}%
\bibitem [{\citenamefont {Schmidt}\ \emph {et~al.}(2019)\citenamefont
  {Schmidt}, \citenamefont {Klein},\ and\ \citenamefont
  {Cancissu~Araujo}}]{Wavepacket3}%
  \BibitemOpen
  \bibfield  {author} {\bibinfo {author} {\bibfnamefont {B.}~\bibnamefont
  {Schmidt}}, \bibinfo {author} {\bibfnamefont {R.}~\bibnamefont {Klein}}, \
  and\ \bibinfo {author} {\bibfnamefont {L.}~\bibnamefont {Cancissu~Araujo}},\
  }\href {\doibase https://doi.org/10.1002/jcc.26045} {\bibfield  {journal}
  {\bibinfo  {journal} {Journal of Computational Chemistry}\ }\textbf {\bibinfo
  {volume} {40}},\ \bibinfo {pages} {2677} (\bibinfo {year} {2019})},\ \Eprint
  {http://arxiv.org/abs/https://onlinelibrary.wiley.com/doi/pdf/10.1002/jcc.26045}
  {https://onlinelibrary.wiley.com/doi/pdf/10.1002/jcc.26045} \BibitemShut
  {NoStop}%
\bibitem [{\citenamefont {{Wolfram Research{,} Inc.}}()}]{Mathematica}%
  \BibitemOpen
  \bibfield  {author} {\bibinfo {author} {\bibnamefont {{Wolfram Research{,}
  Inc.}}},\ }\href {https://www.wolfram.com/mathematica} {\enquote {\bibinfo
  {title} {Mathematica, {V}ersion 12.2},}\ }\bibinfo {note} {Champaign, IL,
  2020}\BibitemShut {NoStop}%
\bibitem [{\citenamefont {Cohen-Tannoudji}\ \emph {et~al.}(1977)\citenamefont
  {Cohen-Tannoudji}, \citenamefont {Diu},\ and\ \citenamefont
  {Lalo{\"e}}}]{cohen1977}%
  \BibitemOpen
  \bibfield  {author} {\bibinfo {author} {\bibfnamefont {C.}~\bibnamefont
  {Cohen-Tannoudji}}, \bibinfo {author} {\bibfnamefont {B.}~\bibnamefont
  {Diu}}, \ and\ \bibinfo {author} {\bibfnamefont {F.}~\bibnamefont
  {Lalo{\"e}}},\ }\href {https://books.google.de/books?id=CnkfAQAAMAAJ} {\emph
  {\bibinfo {title} {Quantum mechanics vol 1}}},\ Quantum Mechanics\ (\bibinfo
  {publisher} {Wiley},\ \bibinfo {year} {1977})\BibitemShut {NoStop}%
\bibitem [{\citenamefont {Sakurai}\ and\ \citenamefont
  {Tuan}(1994)}]{Sakurai1994}%
  \BibitemOpen
  \bibfield  {author} {\bibinfo {author} {\bibfnamefont {J.~J.}\ \bibnamefont
  {Sakurai}}\ and\ \bibinfo {author} {\bibfnamefont {S.~F.~E.}\ \bibnamefont
  {Tuan}},\ }\href@noop {} {\emph {\bibinfo {title} {{Modern Quantum Mechanics
  - Revised Edition}}}}\ (\bibinfo  {publisher} {Addison-Wesley Publishing
  Company},\ \bibinfo {year} {1994})\BibitemShut {NoStop}%
\bibitem [{\citenamefont {van~de Meerakker}\ \emph {et~al.}(2012)\citenamefont
  {van~de Meerakker}, \citenamefont {Bethlem}, \citenamefont {Vanhaecke},\ and\
  \citenamefont {Meijer}}]{meijer_review2012}%
  \BibitemOpen
  \bibfield  {author} {\bibinfo {author} {\bibfnamefont {S.~Y.~T.}\
  \bibnamefont {van~de Meerakker}}, \bibinfo {author} {\bibfnamefont {H.~L.}\
  \bibnamefont {Bethlem}}, \bibinfo {author} {\bibfnamefont {N.}~\bibnamefont
  {Vanhaecke}}, \ and\ \bibinfo {author} {\bibfnamefont {G.}~\bibnamefont
  {Meijer}},\ }\href {\doibase https://doi.org/10.1021/cr200349r} {\bibfield
  {journal} {\bibinfo  {journal} {Chem. Rev.}\ }\textbf {\bibinfo {volume}
  {112}},\ \bibinfo {pages} {4828} (\bibinfo {year} {2012})}\BibitemShut
  {NoStop}%
\bibitem [{\citenamefont {Reber{\v s}ek}\ and\ \citenamefont {Miklav{\v c}i{\v
  c}}(2011)}]{ns_Switching2}%
  \BibitemOpen
  \bibfield  {author} {\bibinfo {author} {\bibfnamefont {M.}~\bibnamefont
  {Reber{\v s}ek}}\ and\ \bibinfo {author} {\bibfnamefont {P.~D.}\ \bibnamefont
  {Miklav{\v c}i{\v c}}},\ }\href {\doibase 10.1080/00051144.2011.11828399}
  {\bibfield  {journal} {\bibinfo  {journal} {Automatika}\ }\textbf {\bibinfo
  {volume} {52}},\ \bibinfo {pages} {12} (\bibinfo {year} {2011})},\ \Eprint
  {http://arxiv.org/abs/https://doi.org/10.1080/00051144.2011.11828399}
  {https://doi.org/10.1080/00051144.2011.11828399} \BibitemShut {NoStop}%
\bibitem [{\citenamefont {Lee}\ \emph {et~al.}(2019)\citenamefont {Lee},
  \citenamefont {Shimizu}, \citenamefont {Funakubo}, \citenamefont {Imai},
  \citenamefont {Sakata}, \citenamefont {Hwang}, \citenamefont {Kim},
  \citenamefont {Yoon}, \citenamefont {Dai}, \citenamefont {Chen},
  \citenamefont {Lee},\ and\ \citenamefont {Jo}}]{ns_Switching1}%
  \BibitemOpen
  \bibfield  {author} {\bibinfo {author} {\bibfnamefont {H.~J.}\ \bibnamefont
  {Lee}}, \bibinfo {author} {\bibfnamefont {T.}~\bibnamefont {Shimizu}},
  \bibinfo {author} {\bibfnamefont {H.}~\bibnamefont {Funakubo}}, \bibinfo
  {author} {\bibfnamefont {Y.}~\bibnamefont {Imai}}, \bibinfo {author}
  {\bibfnamefont {O.}~\bibnamefont {Sakata}}, \bibinfo {author} {\bibfnamefont
  {S.~H.}\ \bibnamefont {Hwang}}, \bibinfo {author} {\bibfnamefont {T.~Y.}\
  \bibnamefont {Kim}}, \bibinfo {author} {\bibfnamefont {C.}~\bibnamefont
  {Yoon}}, \bibinfo {author} {\bibfnamefont {C.}~\bibnamefont {Dai}}, \bibinfo
  {author} {\bibfnamefont {L.~Q.}\ \bibnamefont {Chen}}, \bibinfo {author}
  {\bibfnamefont {S.~Y.}\ \bibnamefont {Lee}}, \ and\ \bibinfo {author}
  {\bibfnamefont {J.~Y.}\ \bibnamefont {Jo}},\ }\href {\doibase
  10.1103/PhysRevLett.123.217601} {\bibfield  {journal} {\bibinfo  {journal}
  {Phys. Rev. Lett.}\ }\textbf {\bibinfo {volume} {123}},\ \bibinfo {pages}
  {217601} (\bibinfo {year} {2019})}\BibitemShut {NoStop}%
\bibitem [{\citenamefont {Gao}\ \emph {et~al.}(2008)\citenamefont {Gao},
  \citenamefont {Drake}, \citenamefont {Chen},\ and\ \citenamefont
  {DeCamp}}]{Gao2008}%
  \BibitemOpen
  \bibfield  {author} {\bibinfo {author} {\bibfnamefont {Y.}~\bibnamefont
  {Gao}}, \bibinfo {author} {\bibfnamefont {T.}~\bibnamefont {Drake}}, \bibinfo
  {author} {\bibfnamefont {Z.}~\bibnamefont {Chen}}, \ and\ \bibinfo {author}
  {\bibfnamefont {M.~F.}\ \bibnamefont {DeCamp}},\ }\href {\doibase
  10.1364/OL.33.002776} {\bibfield  {journal} {\bibinfo  {journal} {Opt.
  Lett.}\ }\textbf {\bibinfo {volume} {33}},\ \bibinfo {pages} {2776} (\bibinfo
  {year} {2008})}\BibitemShut {NoStop}%
\bibitem [{\citenamefont {Arkhipov}\ \emph {et~al.}(2017)\citenamefont
  {Arkhipov}, \citenamefont {Arkhipov}, \citenamefont {Pakhomov}, \citenamefont
  {Babushkin}, \citenamefont {Demircan}, \citenamefont {Morgner},\ and\
  \citenamefont {Rosanov}}]{Arkhipov17}%
  \BibitemOpen
  \bibfield  {author} {\bibinfo {author} {\bibfnamefont {M.~V.}\ \bibnamefont
  {Arkhipov}}, \bibinfo {author} {\bibfnamefont {R.~M.}\ \bibnamefont
  {Arkhipov}}, \bibinfo {author} {\bibfnamefont {A.~V.}\ \bibnamefont
  {Pakhomov}}, \bibinfo {author} {\bibfnamefont {I.~V.}\ \bibnamefont
  {Babushkin}}, \bibinfo {author} {\bibfnamefont {A.}~\bibnamefont {Demircan}},
  \bibinfo {author} {\bibfnamefont {U.}~\bibnamefont {Morgner}}, \ and\
  \bibinfo {author} {\bibfnamefont {N.~N.}\ \bibnamefont {Rosanov}},\ }\href
  {\doibase 10.1364/OL.42.002189} {\bibfield  {journal} {\bibinfo  {journal}
  {Opt. Lett.}\ }\textbf {\bibinfo {volume} {42}},\ \bibinfo {pages} {2189}
  (\bibinfo {year} {2017})}\BibitemShut {NoStop}%
\bibitem [{\citenamefont {Wu}\ and\ \citenamefont {Meyer-ter
  Vehn}(2012)}]{Wu2012}%
  \BibitemOpen
  \bibfield  {author} {\bibinfo {author} {\bibfnamefont {H.-C.}\ \bibnamefont
  {Wu}}\ and\ \bibinfo {author} {\bibfnamefont {J.}~\bibnamefont {Meyer-ter
  Vehn}},\ }\href@noop {} {\bibfield  {journal} {\bibinfo  {journal} {Nature
  Photonics}\ }\textbf {\bibinfo {volume} {6}},\ \bibinfo {pages} {304}
  (\bibinfo {year} {2012})}\BibitemShut {NoStop}%
\bibitem [{\citenamefont {Antonov}\ \emph {et~al.}(2021)\citenamefont
  {Antonov}, \citenamefont {Stollenwerk}, \citenamefont {Venkataramanababu},
  \citenamefont {de~Lima~Batista}, \citenamefont {de~Oliveira-Filho},\ and\
  \citenamefont {Odom}}]{SuperRotors2021}%
  \BibitemOpen
  \bibfield  {author} {\bibinfo {author} {\bibfnamefont {I.~O.}\ \bibnamefont
  {Antonov}}, \bibinfo {author} {\bibfnamefont {P.~R.}\ \bibnamefont
  {Stollenwerk}}, \bibinfo {author} {\bibfnamefont {S.}~\bibnamefont
  {Venkataramanababu}}, \bibinfo {author} {\bibfnamefont {A.~P.}\ \bibnamefont
  {de~Lima~Batista}}, \bibinfo {author} {\bibfnamefont {A.~G.~S.}\ \bibnamefont
  {de~Oliveira-Filho}}, \ and\ \bibinfo {author} {\bibfnamefont {B.~C.}\
  \bibnamefont {Odom}},\ }\href {\doibase 10.1038/s41467-021-22342-6}
  {\bibfield  {journal} {\bibinfo  {journal} {Nature Communications}\ }\textbf
  {\bibinfo {volume} {12}},\ \bibinfo {pages} {2201} (\bibinfo {year}
  {2021})}\BibitemShut {NoStop}%
\end{thebibliography}%

\end{document}